\documentclass[aip,pof,reprint,eqsecnum]{revtex4-1}

\usepackage{natbib}
\usepackage{amssymb,amsmath}

\usepackage{epsfig}
\usepackage{bm}
\usepackage{color}
\usepackage{graphicx}
\usepackage{hyperref}

\newcommand{\wn}{\text{wn}}
\newcommand\beq{\begin{equation}}
\newcommand\eeq{\end{equation}}
\newcommand\beqa{\begin{eqnarray}}
\newcommand\eeqa{\end{eqnarray}}
\def\bal#1\eal{\begin{align}#1\end{align}}
\newcommand{\nn}{\nonumber\\}

\newcommand{\width}{5.2cm}
\newcommand{\widthone}{5.6cm}
\newcommand{\widthtwocm}{5.cm}
\newcommand{\widththree}{6.1cm}
\newcommand{\Tt}{T_t}
\newcommand{\Tr}{T_r}
\newcommand{\zt}{\xi_t}
\newcommand{\zr}{\xi_r}
\newcommand{\at}{\widetilde{\alpha}}
\newcommand{\bt}{\widetilde{\beta}}
\newcommand{\al}{\alpha}
\newcommand{\cc}{\mathbf{v}}
\newcommand{\ww}{\bm{\omega}}
\newcommand{\q}{\kappa}

\newcommand{\qq}[1]{}

\begin{document}

\title{Steady state in a gas of inelastic rough spheres heated by a uniform stochastic force}
\author{Francisco Vega Reyes}\email{fvega@unex.es}
\homepage{http://www.unex.es/eweb/fisteor/fran/fvega.html}
\author{Andr\'es Santos} \email{andres@unex.es}
\homepage{http://www.unex.es/eweb/fisteor/andres/Cvitae/}
\affiliation{Departamento de F\'isica and Instituto de Computaci\'on Cient\'ifica Avanzada (ICCAEx), Universidad de Extremadura, 06071 Badajoz, Spain}


\begin{abstract}
We study here the steady state attained in a granular gas of inelastic rough spheres that is subject to a spatially uniform random volume force. The stochastic force has the form of the so-called white noise and acts by adding impulse to the particle translational  velocities. We work out an analytical solution of the corresponding velocity distribution function from a Sonine polynomial expansion that displays energy non-equipartition between the translational and rotational modes, translational and rotational kurtoses, and  translational-rotational velocity correlations. By comparison with a numerical solution of the Boltzmann kinetic equation (by means of the Direct Simulation Monte Carlo  method) we show that our analytical solution provides a good description that is quantitatively very accurate in certain ranges of inelasticity and roughness.  We also find three important features that make the forced granular gas steady state very different from the homogeneous cooling state (attained by an unforced granular gas). First,  the marginal velocity distributions are always close to a Maxwellian. Second, there is a continuous transition to the purely smooth limit (where the effects of particle rotations are ignored). And third, the angular translational-rotational velocity correlations show a preference for a quasiperpendicular mutual orientation (which is called ``lifted-tennis-ball'' behavior).
\end{abstract}

\date{\today}
\maketitle

\section{Introduction}

Systems composed of a large number of particles with a characteristic length bigger than about $1~\mu\mathrm{m}$ (\textit{granular} systems) are ubiquitous in nature\cite{dG99,B54} and applications.\cite{JNB96a} Therefore, the study of the transport processes in this kind of systems (generically called \textit{granular matter}) has been of interest and subject of systematic study for a long time (see, for instance, the work by Reynolds\cite{R85}). The most generic property of granular matter is the non-conservation of kinetic energy after contact between particles, and hence the term \textit{inelastic} particles.\cite{dG99,JNB96a}

Attending to the nature of the mechanical interaction among grains, granular matter may be classified into two groups: \textit{wet} granular matter (grains tend to stick together after contact and they are usually called \textit{cohesive powders}, since they usually predominate in the smaller range of grain sizes\cite{CVPRW99}) and \textit{dry} granular matter (grains do no not show this sticky behavior). Of course, both types of granular systems are {equally} relevant in industry and technology.\cite{dG99,R90,JNB96a,G03} Moreover, the finding of a meaningful theoretical description also poses fundamental challenges for researchers of a number of fields.\cite{dG99} Since the nature of particle contacts essentially determines the granular dynamics, theoretical developments frequently study granular systems as composed by only dry or only wet particles (or, at its most elaborate, simple combinations of both).

Let us {focus} on dry granular matter. Here, we can find persistent contacts as well, but they are always due to a high  fraction of the volume occupied by grains. When this happens, available space limitations result in lasting contacts. As a consequence, mechanical force transmission throughout the system occurs;\cite{MB05,GG05} i.e., one particle may be in contact with not just one another particle but a set of them, thus propagating mechanical information to large parts of the system.\cite{BISE13} Correlations in this case are obviously strong, much in the same way as in atomic solid matter.\cite{GG05} On the contrary, in less dense systems, particles are less correlated and contacts among three or more grains tend to be statistically rare events; i.e., contacts typically become only \emph{binary}.\cite{G03} At the same time, in the low density regime, those contacts between pairs of dry grains tend to be very short in time {(as compared with the characteristic mean free time)}, and hence the term \textit{inelastic collision}.\cite{BP04}  Binary and instantaneous {collisions} are also characteristic features of low density molecular gases, for which the so-called \textit{molecular chaos} assumption (\textit{Stosszahlansatz}) is {commonly} used;\cite{CC70,BDS97} i.e., colliding particle velocities are supposed to be uncorrelated. In those circumstances, the Boltzmann equation or, at low but finite density, the Enskog equation (properly modified in order to take into account inelasticity in the collisions\cite{BP04,BDS97}) applies. While inelastic collisions may originate velocity correlations,\cite{SM01}  these effects tend to be more important at higher densities\cite{PEU02} and thus they can usually be ignored for low densities.\cite{G03,BDKS98,BP04} Since the theoretical approach to both the high and low density limits for granular matter is radically different\cite{dG99} (actually, just in the same way as it is for conventional matter\cite{CL00}), it is convenient  to focus on just one density regime. In this case we are interested in the low density limit, the so-called \textit{granular gases}.\cite{D01}

Given that most granular particles do not have a regular shape and differ in size and mass,\cite{GDH07} granular collisions (even if short in time) can be very complex {events}.\cite{C00a,KH04} It is thus important that granular collision models can accordingly capture as many features as possible, {without otherwise compromising their tractability}.  For instance, {the coefficient of normal restitution} (that characterizes translational kinetic energy loss upon collisions) depends in general on the impact velocity\cite{BP04,KH04} (a viscoelastic collision model may be used to take this into account\cite{BSSP04}). Nevertheless, it has been found that an important set of the main features at a macroscopic level (i.e., average field behavior) can be described with models that ignore the effects of impact velocity on the restitution properties.\cite{G03} In an effort to simplify the collision process, one can sequentially describe it as (i) an \textit{impact} between both colliding particles, after which they suffer a mechanical impulse that produces a linear momentum transfer; (ii) a \textit{friction} process, tangent to the contact interface, generating an angular momentum transfer; and (iii) a \textit{sliding} process, that describes the relative translational displacement of the colliding particles in the direction of the contact interface. All these three features were captured in previous research, that includes experimental work (see for instance Ref.\ \onlinecite{FLCA94}), by using a collision model based on three collisional constant parameters: {a coefficient of normal restitution, a coefficient of tangential restitution, and a friction coefficient}.

In spite of the intrinsic polydispersity of granular systems in nature,\cite{B54} a variety of generic granular dynamics phenomena can be described by means of kinetic theory for a monodisperse granular gas.  In a granular gas of identical particles,  the relevance of collision models based on constant coefficients of restitution is exemplified by the correct description that the simple \textit{inelastic smooth hard-sphere} model gives of the clustering instability of the so-called homogeneous cooling state (HCS), i.e., an unforced state with uniform hydrodynamic quantities and a time decaying granular temperature.\cite{GZ93,L14,MZB14}  Furthermore, simplifications of the collision model {alleviate} the involved mathematical task of integrating the Boltzmann equation, especially for states more complex than the HCS.\cite{VU09,VSG10} In turn, solutions of the Boltzmann equation have allowed for finding new fundamental properties of granular gas dynamics by means of consistent derivations of the corresponding hydrodynamic theories.\cite{D01} Another important example of this is the \textit{inelastic rough hard-sphere} model, that is able to detect  correlations between the angular and translational velocity components in granular gases.\cite{BPKZ07,RA14} Therefore, it is fruitful to compromise between  simple collision models and more sophisticated theoretical developments at a kinetic theory level, in order to discover new fundamental transport properties of granular gases.

{In its basic implementation,} the rough hard-sphere model neglects the effects of eventual sliding in collisions but is capable of showing a number of relevant results for the dynamics of granular gases. For instance, it has shown that energy non-equipartition between translational and rotational modes occurs,\cite{LHMZ98,SKG10,S11b} and that the distribution function of particle velocities can exhibit strong non-Maxwellian effects.\cite{GNB05,SKS11,VSK14} It has also been determined that strong inelasticity or angular-translational correlations do not necessarily lead to difficulties for reaching a normal (or hydrodynamic) state. \cite{VSK14} This observation has important consequences for the development of hydrodynamic theories for granular gases.\cite{G03,KSG14}. However, {most of} these works addressed only the HCS. Useful and fundamental as this state may be  as a reference for transport theories of granular gases,\cite{GNB05,KSG14} it is hard to observe its spontaneous appearance in nature and also difficult to reproduce in laboratory experiments. In fact, granular gases are most frequently observed under a forcing that keeps the system {steadily} fluidized and at low density.

{In this paper, we study the homogeneous steady state of a granular gas of rough spheres excited by a stochastic volume force that
is modeled by means of the so-called \textit{white noise},\cite {W96,WM96,SBCM98,vNE98,MS00,BC02}  as usually named in statistical physics.
We extend previous theoretical results\cite{SKS11} by taking into account the effects of orientational correlations in particle velocities and, in addition,
by comparing the theoretical predictions with computer simulations. As we will see,  even though the velocity kurtoses and correlations are still not negligible,
the heated granular gas has a more Maxwellian-like distribution function for all values of the roughness parameter than the HCS.\cite{VSK14,RA14}
This justifies the practical validity of our theoretical perturbation approach (truncated Sonine expansion) and its excellent agreement with simulation data. In contrast, the HCS is characterized by large deviations from the Maxwellian distribution for small roughness (especially in what concerns the angular velocity), in which case the perturbation method is valid only semi-quantitatively.\cite{VSK14} Moreover, no singularity exists in the smooth-sphere limit for the heated gas, again in contrast to what happens in the HCS.}

Another important and distinctive feature that we will find is that the forced granular gas shows an average preference  for quasiperpendicular relative orientation of translational and rotational particle velocities, in contrast to the behavior found in the unforced granular gas,\cite{BPKZ07,KBPZ09,RA14,VSK14} where quasiparallel translational and rotational particle velocities {are also present}. This result is of interest for a number of experimental works, where the most common situation is that of an excited granular gas (see, for instance, Ref.\ \onlinecite{LBLG00}).

The structure of this work is as follows. In  Sec.\ \ref{sec2} we write the corresponding kinetic (Boltzmann) equation. By defining adequate time and velocity units, we also derive the  dimensionless counterparts of the Boltzmann equation and its derived moment equations, which are the ones we will work with. In Sec.\ \ref{sec3},  a theoretical solution to the moment equations, by means of a  {Sonine} polynomial expansion around the Maxwellian distribution function, is derived by keeping terms up to order two {(fourth-degree polynomials)}. In this way, the state of the system is solved \emph{at all times}, including the final steady state.
In particular, the theoretical parameters characterizing kurtoses, correlations, and average relative orientation of translational and rotational particle velocities are analyzed. As said before, we will find that excited rough grains do not show the relevance of quasiparallel relative orientation behavior present in the HCS.\cite{BPKZ07,VSK14} Next, in Sec.\ \ref{sec4} we  compare the theoretical results  with an  ``exact'' numerical solution of the Boltzmann equation obtained by means of the Direct Simulation Monte Carlo  (DSMC) method,\cite{B94}  a very good agreement being observed. Finally, the results are briefly discussed in Sec.\ \ref{sec5}.

\section{Boltzmann equation and moment hierarchy}
\label{sec2}

\subsection{Collision rules}
Let us consider a homogeneous dilute granular gas of identical hard spheres with mass $m$, diameter $\sigma$, and moment of inertia $I$, subject to a stochastic volume force $\mathbf{F}^\wn$ (also called \textit{thermostat}) with the properties of a Gaussian white noise.\cite{WM96,W96,SBCM98,vNE98,MS00,GSVP11a} This kind of forcing can model, for example, the energy input to grains immersed in a gas in turbulent flow.\cite{OLDLD04}

We denote the translational and angular (or rotational) particle velocities with $\mathbf{v}$ and $\bm{\omega}$ respectively. Binary collisions are characterized here by two material parameters, namely the coefficient of normal restitution $\alpha$ and the coefficient of tangential restitution $\beta$, {which determine the shrinking of the  normal and tangential component, respectively, of the relative velocity of the two surface points at contact.} They are defined by (see, for instance, Refs.\ \onlinecite{K10a,S10,KSG14})
\begin{equation}
\label{collrule}
\widehat{\bm{\sigma}}\cdot\mathbf{u}'=-\alpha(\widehat{\bm{\sigma}}\cdot\mathbf{u}), \quad \widehat{\bm{\sigma}}\times\mathbf{u}'=-\beta(\widehat{\bm{\sigma}}\times\mathbf{u}).
\end{equation}
{Here, the primes denote post-collisional values, $\widehat{\bm{\sigma}}$ is the unit collision vector  joining the centers of the two colliding spheres (and pointing from the center of  particle 1 to the center of particle 2), and}
\beq
{\mathbf{u}=\mathbf{v}_1-\mathbf{v}_2-{\frac{\sigma}{2}}\widehat{\bm{\sigma}}\times(\bm{\omega}_1+\bm{\omega}_2)}
 \label{3a}
 \eeq
{is the  relative velocity  of the points of the spheres which are in contact during a binary encounter.}

The coefficient of normal restitution $\alpha$ takes values between $0$ (completely inelastic collision) and $1$ (completely elastic collision), while the coefficient of tangential restitution takes values between $-1$ (completely smooth collision, implying that rotational velocities are unchanged) and $1$ (completely rough collision).\cite{CC70,K10a} {Equation \eqref{collrule} supplements the linear and angular momentum conservation laws to yield the  collision rules},\cite{K10a,SKS11,BP04,KSG14}
\beq
\label{1}
 {m \mathbf{v}_{1,2}'=m\mathbf{v}_{1,2} \mp\mathbf{Q},\quad I \bm{\omega}_{1,2}'=I\bm{\omega}_{1,2} -{\frac{\sigma}{2}}\widehat{\bm{\sigma}}\times\mathbf{Q},}
\eeq
{where  the impulse exerted by  particle 1 on particle 2 is given by}
 \beq
{\mathbf{Q}={m\widetilde\alpha}(\widehat{\bm{\sigma}}\cdot\mathbf{u})\widehat{\bm{\sigma}}-m\widetilde\beta\widehat{\bm{\sigma}}
\times(\widehat{\bm{\sigma}}\times\mathbf{u}).}
\label{6}
 \eeq
{Here, we have introduced  the abbreviations}
 \beq
 \label{7}
{ \widetilde\alpha\equiv{\frac{1+\alpha}{2}},\quad \widetilde\beta\equiv{\frac{1+\beta}{2}}{\frac{\kappa}{\kappa+1}},
\quad \kappa\equiv {\frac{4I}{m\sigma^2}}.}
 \eeq

{While collisions (except if $\alpha=1$ and $\beta=\pm 1$) dissipate kinetic energy (translational plus rotational), the homogeneous \emph{stochastic} force $\mathbf{F}^\wn$ injects translational kinetic energy to grains. It has properties of a Gaussian white noise, i.e.,}
\begin{subequations}
\bal
\langle {\bf F}_i^{\wn}(t) \rangle =&{\bf 0},  \\
\langle {\bf F}_i^{\wn}(t) {\bf F}_j^{\wn}(t') \rangle =&\mathsf{I}m^2 \chi_0^2 \delta_{ij}\delta(t-t'),
\label{Fst}
\eal
\end{subequations}
where indexes $i,j$ refer to  particles, $\mathsf{I}$ is the $3\times 3$ unit matrix, and $\chi_0^2$  measures the characteristic strength of the stochastic force.

\subsection{Boltzmann equation}
In homogeneous states,  the  Boltzmann equation corresponding to the stochastic external force $\mathbf{F}^\wn$ becomes\cite{vNE98,MS00}
\beq
\frac{\partial f(\mathbf{v},\bm{\omega};t)}{\partial t}-\frac{\chi_0^2}{2}\left(\frac{\partial}{\partial \cc}\right)^2f(\mathbf{v},\bm{\omega};t)={J[\cc,\ww|f(t)]},
\label{BE}
\eeq
where $f(\mathbf{v},\bm{\omega};t)$ is the velocity distribution function and $J[\cc,\ww|f]$ is the collision integral in the (inelastic) Boltzmann equation for rough spheres, {which accounts for the collision rules \eqref{1} and \eqref{6}.}

{Given a certain function $A(\cc,\ww)$, we define its average as}
\beq
\langle A(t)\rangle=\frac{1}{n}\int d\cc\int d\ww\, A(\cc,\ww)f(\cc,\ww;t), \quad n=\int d\cc\int d\ww\, f(\cc,\ww;t),
\eeq
where
$n$ is the number density. By Galilean invariance, we choose $\langle \cc\rangle=\mathbf{0}$. Moreover, we assume isotropic states, so that $\langle\ww\rangle=\mathbf{0}$.\cite{KSG14} Thus, the basic quantities are the translational ($\Tt$), rotational ($\Tr$), and total ($T$) granular temperatures:
\beq
{\Tt={\frac{m}{3}}\langle v^2\rangle,\quad \Tr={\frac{I}{3}}\langle \omega^2\rangle,\quad \quad T=\frac{\Tt+\Tr}{2}=\Tt\frac{1+\theta}{2}.}
\label{Tt,Tr}
\eeq
where the temperature ratio $\theta\equiv \Tr/\Tt$ is an important quantity in this system because, as we will see, its steady-state value is independent of the level of forcing ($\chi_0^2$).

The evolution equations for the granular temperatures $\Tt$ and $\Tr$ are obtained by just multiplying both sides of Eq.\ \eqref{BE} by the particle translational and rotational kinetic energies, respectively, and integrating over all velocity values. This gives
\beq
\partial_t\Tt-{m\chi_0^2}=-\zt \Tt,\quad {\partial_t\Tr=-\zr \Tr}, \quad \partial_t T-\frac{m\chi_0^2}{2}=-\zeta  T
\label{Tidt}
\eeq
Here, the translational ($\zt$) and rotational ($\zr$) {production rates and the cooling rate ($\zeta$)} are defined by the collision integrals\cite{SKS11,KSG14}
\begin{subequations}
\begin{align}
  \label{coolingrates}
  & \zt=-{\frac{m}{3n\Tt}}\int d\mathbf{v}\int d\bm{\omega}\,v^2{J[\mathbf{v},\bm{\omega}| f]},
  \quad
  \zr=-{\frac{I}{3n\Tr}}\int d\mathbf{v}\int d\bm{\omega}\,\omega^2{J[\mathbf{v},\bm{\omega}| f]}, \\
& \zeta=\frac{\zt \Tt+\zr \Tr}{2T}=\frac{\zt+\zr{\theta}}{1+\theta}.
\end{align}
\end{subequations}

\subsection{Reduced variables and velocity moments}
For convenience, let us rewrite the Boltzmann equation \eqref{BE} in dimensionless form. For this, we first define the reduced velocity distribution function
\beq
\phi(\mathbf{c},\mathbf{w};\tau)\equiv \frac{1}{n}\left(\frac{4T_t(t)T_r(t)}{m I}\right)^{3/2}f(\mathbf{v}, \bm{\omega};t), \label{phi}
\eeq
where the reduced particle velocities are
\beq
\mathbf{c}(t)\equiv\frac{\mathbf{v}}{\sqrt{2T_t(t)/m}}, \quad
\mathbf{w}(t)\equiv\frac{\bm{\omega}}{\sqrt{2T_r(t)/I}}.
\label{cw}
\eeq
{In addition, time is measured by the parameter $\tau$ defined by}
\beq
{\tau(t)=\int_0^t  dt'\,\nu(t'),\quad \nu(t)=2n\sigma^2 \sqrt{\pi T_t(t)/m},}
\eeq
where {$\nu(t)$ is the (time-dependent) collision frequency. Thus, $\tau(t)$ measures the accumulated number of collisions per particle up to time $t$.} With this time scale, we can measure relaxation times to the steady state that are physically meaningful, since they coincide with the aging times to hydrodynamics.\cite{VSK14}

The  moments of the reduced distribution function $\phi(\mathbf{c},\mathbf{w};\tau)$ are
\beq
M_{pq}^{(r)}(\tau)\equiv \langle c^pw^q
(\mathbf{c}\cdot\mathbf{w})^r\rangle=\int d\mathbf{c}\int d\mathbf{w}\,c^pw^q
(\mathbf{c}\cdot\mathbf{w})^r\phi(\mathbf{c},\mathbf{w};\tau),
\eeq
{with $p,q,r=\text{even}$, by symmetry. Note that $M_{pq}^{(r)}$ is a moment of degree $p+q+2r$ and $M_{0,0}^{(0)}=1$, $M_{2,0}^{(0)}=M_{0,2}^{(0)}=\frac{3}{2}$.}
The dimensionless measure of the noise intensity is defined by
\beq
\gamma(\tau)\equiv \frac{3}{2}\frac{\chi_0^2}{\nu(t)\Tt(t)/m}=\frac{3\chi_0^2}{4\sqrt{\pi}n\sigma^2[\Tt(t)/m]^{3/2}}.
\label{Gamma}
\eeq
Notice that, although  $\chi_0^2$ is a constant, its dimensionless counterpart $\gamma$ is time-dependent  due to the scaling with the temperature $\Tt$. Of course, once a steady state is reached, $\gamma$ becomes constant too.
{Note also that $\gamma^{-2/3}$ can be seen as the translational temperature $\Tt$ in units of a reference temperature $T_{\text{ref}}=m(3\chi_0^2/4\sqrt{\pi}n\sigma^2)^{2/3}$ defined from the white noise parameter $\chi_0^2$.}

Let us finally define the reduced collisional moments
\beq
  \label{mupq}
\mu_{pq}^{(r)}(\tau)\equiv-\int d\mathbf{c}\int d\mathbf{w}\,
c^pw^q
(\mathbf{c}\cdot\mathbf{w})^r
{\mathcal{J}[\mathbf{c},\mathbf{w}|\phi(\tau)]},
\eeq
where
\begin{equation}
  \label{Cop}
\mathcal{J}[\mathbf{c}, \mathbf{w}|\phi(\tau)]=\frac{1}{n\nu(t)}\left[\frac{4T_t(t)T_r(t)}{m I}\right]^{3/2} J[\cc,\ww|f(t)]
\end{equation}
is the dimensionless form of the collision integral. {}From the definitions \eqref{coolingrates} and \eqref{mupq}, it is easy to see that
$\zt=\frac{2}{3}\nu\mu_{20}^{(0)}$, $\zr=\frac{2}{3}\nu\mu_{02}^{(0)}$. The evolution equations for the temperature ratio $\theta(\tau)$ and the reduced noise strength $\gamma(\tau)$ can be obtained from Eq.\ \eqref{Tidt} as
\beq
\frac{\partial \ln\theta(\tau)}{\partial \tau}=\frac{2}{3}\left[\mu_{20}^{(0)}(\tau)-\mu_{02}^{(0)}(\tau)-\gamma(\tau)\right],
\quad
\frac{\partial\ln\gamma(\tau)}{\partial \tau}=\mu_{20}^{(0)}(\tau)-\gamma(\tau).
\label{thetadt}
\eeq
{Note that, since $\cc$ and $\ww$ are scaled each with a different temperature in Eq.\ \eqref{cw}, the collision rules \eqref{1} and \eqref{6}, when expressed in terms of $\mathbf{c}$ and $\mathbf{w}$, depend parametrically on $\theta$.\cite{SKS11} This dependence is obviously transferred to the collisional moments \eqref{mupq}.}

Taking into account the previous dimensionless quantities, {and making use again of the temperature evolution equations \eqref{Tidt}}, one can obtain the dimensionless version of the Boltzmann equation \eqref{BE} as
\bal
\frac{\partial \phi(\mathbf{c},\mathbf{w};\tau)}{\partial\tau}&+\frac{\mu_{20}^{(0)}(\tau)-
\gamma(\tau)}{3}\frac{\partial}{\partial\mathbf{c}}\cdot\left[\mathbf{c}\phi(\mathbf{c},\mathbf{w};\tau)
\right]+\frac{\mu_{02}^{(0)}(\tau)}{3}\frac{\partial}{\partial\mathbf{w}}\cdot\left[\mathbf{w}
\phi(\mathbf{c},\mathbf{w};\tau)\right]\nn
& -\frac{\gamma(\tau)}{6}\left(\frac{\partial}{\partial \mathbf{c}}\right)^2\phi(\mathbf{c},\mathbf{w};\tau)= {\mathcal{J}[\mathbf{c},\mathbf{w}|\phi(\tau)]}.
\label{rBE}
\eal
Multiplying both sides by $c^pw^q
(\mathbf{c}\cdot\mathbf{w})^r$ and integrating over $\mathbf{c}$ and $\mathbf{w}$, we get the moment hierarchy
\bal
\frac{\partial \ln M_{pq}^{(r)}(\tau)}{\partial\tau}=&
\frac{p+r}{3}\left[\mu_{20}^{(0)}(\tau)-\gamma(\tau)\right]+\frac{q+r}{3}{\mu_{02}^{(0)}}(\tau)-
\frac{\mu_{pq}^{(r)}(\tau)}{M_{pq}^{(r)}(\tau)}\nn
&+\frac{\gamma(\tau)}{6}\frac{p(p+1+2r)M_{p-2,q}^{(r)}(\tau)
+r(r-1)M_{p,q+2}^{(r-2)}(\tau)}{M_{pq}^{(r)}(\tau)}.
\label{partialpq}
\eal
This equation must be supplemented with Eq.\ \eqref{thetadt}.
As will be seen later, the steady-state distribution function is independent of the precise value of the stochastic force intensity $\chi_0^2$ or, equivalently, of the initial value $\gamma(0)$, although of course the transient states are dependent on it.

\subsection{Exact Sonine expansion}
{In principle, the reduced distribution function $\phi(\mathbf{c},\mathbf{w})$ would be a function of six velocity components. On the other hand, since we restrict ourselves to isotropic states, $\phi$ is actually a function of three scalar quantities: the norms $c^2$ and $w^2$, and the squared dot product $(\mathbf{c},\mathbf{w})^2$.\cite{SKS11}} Thus, it makes sense to represent $\phi(\mathbf{c},\mathbf{w})$ as an expansion in a complete set $\{\Psi_{jk}^{(\ell)}(\mathbf{c},\mathbf{w})\}$ of orthogonal polynomials,
\beq
\phi(\mathbf{c},\mathbf{w};\tau)=\phi_M(c,w)\sum_{j=0}^\infty\sum_{k=0}^\infty\sum_{\ell=0}^\infty a_{jk}^{(\ell)}(\tau)\Psi_{jk}^{(\ell)}(\mathbf{c},\mathbf{w}),\quad \phi_M(c,w)\equiv \pi^{-3}e^{-c^2-w^2},
\label{phi_sonine}
\eeq
where
\beq
\Psi_{jk}^{(\ell)}(\mathbf{c},\mathbf{w})\equiv L_j^{(2\ell+\frac{1}{2})}(c^2) L_k^{(2\ell+\frac{1}{2})}(w^2)\left(c^2w^2\right)^\ell
P_{2\ell}(\cos\vartheta),\quad \cos\vartheta\equiv \frac{\mathbf{c}\cdot\mathbf{w}}{cw},
\label{sonine}
\eeq
where $L_j^{(2\ell+\frac{1}{2})}(x)$ and $P_{2\ell}(x)$ are the Laguerre and Legendre polynomials, respectively.\cite{AS72}
Defining the scalar product of two arbitrary isotropic (real) functions $\Phi_1(\mathbf{c},\mathbf{w})$ and $\Phi_2(\mathbf{c},\mathbf{w})$ as
\beq
\langle \Phi_1|\Phi_2\rangle\equiv \int d\mathbf{c}\int d\mathbf{w} \,\phi_M(c,w) \Phi_1(\mathbf{c},\mathbf{w})\Phi_2(\mathbf{c},\mathbf{w}),
\eeq
one can obtain the orthogonality relation
\beq
\langle \Psi_{jk}^{(\ell)}|\Psi_{j'k'}^{(\ell')}\rangle=\mathcal{N}_{jk}^{(\ell)}\delta_{j,j'}\delta_{k,k'}\delta_{\ell,\ell'},\quad
\mathcal{N}_{jk}^{(\ell)}\equiv \frac{\Gamma(j+2\ell+\frac{3}{2})\Gamma(k+2\ell+\frac{3}{2})}{(\pi/4)(4\ell+1)j!k!}.
\eeq
As a consequence, the coefficients in the expansion \eqref{phi_sonine} are
\bal
a_{jk}^{(\ell)}=&\frac{\langle\Psi_{jk}^{(\ell)}\rangle}{\mathcal{N}_{jk}^{(\ell)}}\nn
=&\sum_{j_1=0}^j\sum_{k_1=0}^k\sum_{\ell_1=0}^\ell
\frac{(-1)^{j_1+k_1+\ell_1} (\pi/4)j!k!(4\ell+1)(4\ell-2\ell_1-1)!! M_{2(j_1+\ell_1),2(k_1+\ell_1)}^{(2\ell-2\ell_1)}}{j_1!(j-j_1)!\Gamma\left(j_1+2\ell+\frac{3}{2}\right)
k_1!(k-k_1)!\Gamma\left(k_1+2\ell+\frac{3}{2}\right)2^{\ell_1}\ell_1!(2\ell-2\ell_1)!},
\eal
{where in the second step use has been made of the explicit expressions of the Laguerre and Legendre polynomials.\cite{AS72} We observe that $a_{jk}^{(\ell)}$ is a linear combination of the moments $M_{pq}^{(r)}$ with $p,q,r=\text{even}$ and degree $2\ell\leq p+q+2r\leq 2(j+k+2\ell)$.
By normalization, $a_{00}^{(0)}=1$, $a_{10}^{(0)}=a_{01}^{(0)}=0$. Therefore, the first nontrivial coefficients are those associated with moments of  degree four, namely}
\begin{subequations}
\label{cumulants}
\begin{equation}
a_{20}^{(0)}=\frac{4}{15}\langle c^4\rangle-1, \quad a_{02}^{(0)}=\frac{4}{15}\langle w^4\rangle-1,
\label{a20}
\end{equation}
\begin{equation}
 a_{11}^{(0)}=\frac{4}{9}\langle c^2w^2\rangle-1,\quad
{a_{00}^{(1)}=\frac{8}{15}\left[\langle (\mathbf{c}\cdot\mathbf{w})^2\rangle-\frac{1}{3}\langle c^2w^2\rangle\right]}.
\label{b}
\end{equation}
\label{ajkl}
\end{subequations}
{These are the {fourth-degree} \emph{cumulants}, which  measure the basic deviations of the velocity distribution function $\phi(\mathbf{c},\mathbf{w})$ from the (two-temperaure) Maxwellian $\phi_M(c,w)$. While $a_{20}^{(0)}$ and $a_{02}^{(0)}$  measure the \emph{kurtosis} of the (marginal) translational and rotational velocity distribution functions, respectively, the coefficient $a_{11}^{(0)}$ characterizes the \emph{scalar} translational-rotational correlations, i.e., $\langle c^2 w^2\rangle\neq \langle c^2\rangle\langle w^2\rangle$. Finally, $a_{00}^{(1)}$ informs us about the possible existence of \emph{orientational} correlations, i.e., $\langle c^2w^2\cos^2\vartheta\rangle\neq \langle c^2w^2\rangle\langle \cos^2\vartheta\rangle$ and $\langle \cos^2\vartheta\rangle\neq \frac{1}{3}$. To disentangle the latter two effects, let us define the quantities}
\beq
{h\equiv\frac{5}{8}\left[\frac{\langle(\mathbf{c}\cdot\mathbf{w})^2\rangle}{\langle c^2 w^2\rangle\langle\cos^2\vartheta\rangle}-1\right]},\quad
b\equiv \frac{10}{3}\left(\langle \cos^2\vartheta\rangle-\frac{1}{3}\right).
\label{eqb}
\eeq
{The parameters $h$, $b$, $a_{11}^{(0)}$, and $a_{00}^{(1)}$ are related by}
\beq
{h=\frac{1}{16}\frac{25a_{00}^{(1)}-9b\left(1+a_{11}^{(0)}\right)}{\left(1+\frac{9}{10}b\right)\left(1+a_{11}^{(0)}\right)}.}
\label{5.14}
\eeq

In previous works,\cite{BPKZ07,KBPZ09,RA14} the angle $\vartheta$ between translational and rotational particle velocities has been used for measuring orientational correlations in a granular gas of rough spheres, by tracking the quantity $\langle\cos^2\vartheta\rangle$ or, equivalently, $b$.
The use of $b$  has the advantage over $a_{00}^{(1)}$ of being more intuitive (since it  directly refers to the angle between translational and rotational particle velocities). However, it has the disadvantage that, since $\cos\vartheta=\mathbf{c}\cdot\mathbf{w}/cw$ is not a polynomial, the parameter $b$ involves an infinite number of coefficients in the expansion \eqref{phi_sonine}. More specifically,
it is easy to find
\beq
b=\sum_{j=0}^\infty\sum_{k=0}^\infty a_{jk}^{(1)}.
\label{5.13}
\eeq
Analogously, the quantity $h$ also involves an infinite number of coefficients. For this reason, we prefer instead to use the single coefficient $a_{00}^{(1)}$ for measuring the orientational correlations in a granular gas of rough spheres.\cite{VSK14}

{Taking into account the general moment hierarchy \eqref{partialpq}, the evolution equations for the fourth-degree cumulants are}
\begin{subequations}
\label{evocum}
\beq
\frac{\partial \ln\left[1+a_{20}^{(0)}(\tau)\right]}{\partial \tau} =\frac{4}{3}\mu_{20}^{(0)}(\tau)-\frac{4}{3}\gamma(\tau)-\frac{4}{15}\frac{\mu_{40}^{(0)}(\tau)-5\gamma(\tau)}{1+a_{20}^{(0)}(\tau)},
\label{evoa20}
\eeq
\beq
\frac{\partial \ln\left[1+a_{02}^{(0)}(\tau)\right]}{\partial \tau} =\frac{4}{3}\mu_{02}^{(0)}(\tau)-\frac{4}{15}\frac{\mu_{04}^{(0)}(\tau)}{1+a_{02}^{(0)}(\tau)},
\label{evoa02}
\eeq
\beq
\frac{\partial\ln\left[1+a_{11}^{(0)}(\tau)\right] }{\partial \tau } =\frac{2}{3}\mu_{20}^{(0)}(\tau)+\frac{2}{3}\mu_{02}^{(0)}(\tau)-\frac{2}{3}\gamma(\tau)-
\frac{4}{9}\frac{\mu_{22}^{(0)}(\tau)-\frac{3}{2}\gamma(\tau)}{1+a_{11}^{(0)}(\tau)},
\label{evoa11}
\eeq
\begin{equation}
\frac{\partial \ln\left[1+a_{11}^{(0)}(\tau)+\frac{5}{2}a_{00}^{(1)}(\tau)\right] }{\partial \tau} =\frac{2}{3}\mu_{20}^{(0)}(\tau)+\frac{2}{3}\mu_{02}^{(0)}(\tau)-\frac{2}{3}\gamma(\tau)
-\frac{4}{3}\frac{\mu_{00}^{(2)}(\tau)-\frac{1}{2}\gamma(\tau)}{1+a_{11}^{(0)}(\tau)+\frac{5}{2}a_{00}^{(1)}(\tau)}.
\label{evob}
\end{equation}
\end{subequations}

{In the steady state ($\partial_\tau\to 0$), Eqs.\ \eqref{Tidt} and \eqref{evocum} imply the conditions}
\begin{subequations}
\label{cumss}
\beq
\gamma=\mu_{20}^{(0)}, \quad \mu_{02}^{(0)}=0,
\label{5.8}
\eeq
\beq
 \frac{1}{5}\mu_{40}^{(0)}=\frac{2}{3}\mu_{22}^{(0)}=2\mu_{00}^{(2)}=\mu_{20}^{(0)},  \quad \mu_{04}^{(0)}=0.
\label{5.12}
\eeq
\end{subequations}

{Before closing this section, and for further use, let us define the \emph{marginal} distributions $\phi_{c}(c)$, $\phi_{w}(w)$, $\phi_{cw}(c^2w^2)$, $\phi_{\mathbf{c}\cdot\mathbf{w}}((\mathbf{c}\cdot\mathbf{w})^2)$, and $\phi_\vartheta(\cos^2\vartheta)$ from the \emph{joint} velocity distribution function $\phi(\mathbf{c},\mathbf{w})$ as}
\begin{subequations}
\label{6.1+6.2+6.5}
\beq
\phi_{c}(c)=4\pi c^2\int d\mathbf{w}\, \phi(\mathbf{c},\mathbf{w}),\quad \phi_{w}(w)=4\pi w^2\int d\mathbf{c}\, \phi(\mathbf{c},\mathbf{w}),
\eeq
\beq
\phi_{cw}(c^2w^2)=\int d\mathbf{c}'\int d\mathbf{w}'\,\delta(c'^2w'^2-c^2w^2)\phi(\mathbf{c}',\mathbf{w}'),
\label{6.1}
\eeq
\beq
\phi_{\mathbf{c}\cdot\mathbf{w}}((\mathbf{c}\cdot\mathbf{w})^2)=\int d\mathbf{c}'\int d\mathbf{w}'\,\delta((\mathbf{c}'\cdot\mathbf{w}')^2-(\mathbf{c}\cdot\mathbf{w})^2)\phi(\mathbf{c}',\mathbf{w}'),
\label{6.2}
\eeq
\beq
\phi_\vartheta(\cos^2\vartheta)=\int d\mathbf{c}'\int d\mathbf{w}'\,\delta(\cos^2\vartheta'-\cos^2\vartheta)\phi(\mathbf{c}',\mathbf{w}').
\label{6.5}
\eeq
\end{subequations}
{The less conventional marginal distribution functions \eqref{6.1}--\eqref{6.5} are directly related to the scalar and orientational correlations between the translational and rotational particle velocities. More specifically,}
\beq
\label{c2w2...}
\langle c^2w^2\rangle=\int_0^\infty dx\, x \phi_{cw}(x),\quad \langle (\mathbf{c}\cdot\mathbf{w})^2\rangle=
\int_0^\infty dx\, x \phi_{\mathbf{c}\cdot\mathbf{w}}(x),\quad
\langle \cos^2\vartheta\rangle =\int_0^1 dx\, x \phi_{\vartheta}(x).
\eeq

\section{Approximate solution from a truncated Sonine expansion}
\label{sec3}

{All of the equations in Sec.\ \ref{sec2} are \emph{formally} exact. On the other hand, the collisional moments $\mu_{20}^{(0)}$, $\mu_{02}^{(0)}$, $\mu_{40}^{(0)}$, $\mu_{04}^{(0)}$, $\mu_{22}^{0)}$, and $\mu_{00}^{(2)}$, defined by the general expression \eqref{mupq}, are functionals of $\phi(\mathbf{c},\mathbf{w})$, and thus they depend on the whole set of expansion coefficient $\{a_{jk}^{(\ell)}\}$. They also depend on the temperature ratio $\theta$ through the collision rules. Therefore, Eqs.\ \eqref{thetadt} and \eqref{evocum} do not make a closed set of equations.}

Nevertheless, the expansion \eqref{phi_sonine} is especially useful for states with a distribution function close to the Maxwellian $\phi_M(c,w)$, in which case an approximate \emph{closure} can be applied. In a previous work,\cite{VSK14} we used the  expansion \eqref{phi_sonine} for describing the HCS distribution function. We found that the HCS is very far from the Maxwellian for a certain range of values of the roughness parameter $\beta$, thus limiting the practical validity of this kind of theoretical description for the unforced gas. However, we can expect that the white noise thermostat keeps the distribution function closer to the Maxwellian, {in analogy with what happens in the smooth-sphere case.\cite{MS00}} If so, a better accuracy of the expansion \eqref{phi_sonine}  can be expected in the present case. We will confirm this property in Sec.\ \ref{sec4}.

\subsection{Maxwellian approximation}
{The simplest approximation consists in neglecting all the cumulants $a_{jk}^{(\ell)}$, $j+k+2\ell\geq 2$, in Eq.\ \eqref{phi_sonine}, i.e., $\phi(\mathbf{c},\mathbf{w})\simeq\phi_M(c,w)$. This Maxwellian approximation is of course unable to account for the evolution and steady-state values of the cumulants. However, it can capture the main features of the temperature ratio $\theta(\tau)$ and the reduced noise intensity $\gamma(\tau)$. When $\phi(\mathbf{c},\mathbf{w})\simeq\phi_M(c,w)$ is inserted into Eq.\ \eqref{mupq}, one obtains}
\begin{subequations}
\beq
\mu_{20,M}^{(0)}=1-\alpha^2+\frac{\kappa(1+\beta)}{(1+\kappa)^2}\left[2+\kappa(1-\beta)-\theta(1+\beta)\right], \label{mu20M}
\eeq
\beq
\mu_{02,M}^{(0)}=\frac{\kappa(1+\beta)}{(1+\kappa)^2}\left[2+\kappa^{-1}(1-\beta)-\theta^{-1}(1+\beta)\right]. \label{mu02M}
\eeq
\end{subequations}
{Using these expressions in \eqref{thetadt}, one gets a closed set of two coupled nonlinear differential equations that can be numerically solved for arbitrary initial values $\theta(0)$ and $\gamma(0)$. Regardless of the values of $\theta(0)$ and $\gamma(0)$, common steady-state values are reached after a few collisions per particle. They are explicitly obtained from Eq.\ \eqref{5.8} as}
\beq
\label{thetaMax}
\theta_M=\frac{1+\beta}{2+\kappa^{-1}(1-\beta)},\quad
\gamma_M=1-\alpha^2+\frac{2(1-\beta^2)}{2+\kappa^{-1}(1-\beta)}.
\eeq
{It is interesting to note that the Maxwellian prediction for the steady-state value of the temperature ratio is independent of the coefficient of normal restitution $\alpha$. Moreover, $\theta_M$ is always smaller than unity (i.e., $T_r<T_t$), except in the limit of completely rough spheres ($\beta=1$), where $\theta_M=1$.
In the case of the reduced noise intensity, we can observe that $\gamma_M$ monotonically decreases as $\alpha$ increases at fixed $\beta$, as a consequence of a progressively smaller cooling effect. Because of the same reason, $\gamma_M$ presents a non-monotonic behavior with respect to $\beta$ at fixed $\alpha$, reaching a maximum value $\gamma_{M,\max}=1-\alpha^2+4\kappa(1+2\kappa-2\sqrt{\kappa+\kappa^2})$ at $\beta=1-2(\sqrt{\kappa+\kappa^2}-\kappa)$.
However, as we will see in Sec.\ \ref{sec4}, $\theta$ actually depends on $\alpha$  but much more weakly than it depends on $\beta$. Also, $\theta$ can reach values slightly larger than unity if $\beta$ is very close to $1$ and $\alpha$ is small enough. In any case, as shown later, the Maxwellian expressions \eqref{thetaMax} are indeed rather accurate.}

{The marginal distribution functions defined by \eqref{6.1+6.2+6.5} adopt simple forms in the Maxwellian approximation,}
\begin{subequations}
\beq
\phi_{c,M}(c)=\frac{4}{\sqrt{\pi}}c^2 e^{-c^2},\quad \phi_{w,M}(c)=\frac{4}{\sqrt{\pi}}w^2 e^{-w^2},
\eeq
\beq
\label{6.3}
\phi_{cw,M}(c^2w^2)=\frac{8cw}{\pi}K_0(2cw),\quad
\phi_{\mathbf{c}\cdot\mathbf{w},M}((\mathbf{c}\cdot\mathbf{w})^2)=\frac{4}{\pi}K_1(2|\mathbf{c}\cdot\mathbf{w}|),\quad
\phi_{\vartheta,M}(\cos^2\vartheta)=\frac{1}{2|\cos\vartheta|},
\eeq
\end{subequations}
where $K_n(x)$ is a modified Bessel function of the second kind.\cite{AS72} {Upon deriving \eqref{6.3} use has been made  of the mathematical properties $\delta(c'^2w'^2-x)=\delta(w'-\sqrt{x}/c')/2c'^2w'$,
$\delta((\mathbf{c}'\cdot\mathbf{w}')^2-x)=\delta(|\cos\vartheta|-\sqrt{x}/cw)/2c^2w^2|\cos\vartheta|$, and $\delta(\cos^2\vartheta-x)=\delta(|\cos\vartheta|-\sqrt{x})/2|\cos\vartheta|$, respectively. Apart from $\langle c^4\rangle=\langle w^4\rangle=\frac{15}{4}$,
it is straightforward to check from Eqs.\ \eqref{c2w2...} and \eqref{6.3} that $\langle c^2 w^2\rangle=\frac{9}{4}$, $\langle(\mathbf{c}\cdot\mathbf{w})^2\rangle=\frac{3}{4}$, and $\langle \cos^2\vartheta\rangle=\frac{1}{3}$ in the Maxwellian approximation, as expected.}

\subsection{Truncated Sonine approximation}
{As a much more elaborate approximation that captures the most important non-Maxwellian features of the velocity distribution function, we truncate the exact expansion \eqref{phi_sonine} after $j+k+2\ell=2$, i.e.\cite{VSK14}}
\bal
\frac{\phi(\mathbf{c},\mathbf{w})}{\phi_M(c,w)}\simeq & {1+}a_{20}^{(0)}\Psi_{20}^{(0)}(\mathbf{c},\mathbf{w})+a_{02}^{(0)}\Psi_{02}^{(0)}(\mathbf{c},
\mathbf{w})+a_{11}^{(0)}\Psi_{11}^{(0)}(\mathbf{c},\mathbf{w})+a_{00}^{(1)}\Psi_{00}^{(1)}(\mathbf{c},\mathbf{w})\nn
=&1+{a_{20}^{(0)}}\frac{{15}-20c^2+4c^4}{8}+{a_{02}^{(0)}}\frac{{15}-20w^2+4w^4}{8}+{a_{11}^{(0)}} \frac{\left(3-2c^2\right)\left(3-2w^2\right)}{4}\nn
&+{a_{00}^{(1)}}\frac{3(\mathbf{c}\cdot\mathbf{w})^2-c^2w^2}{2}.
\label{phi2}
\eal
{This approximation differs from that considered in Ref.\ \onlinecite{SKS11} by the addition of the term headed by the coefficient $a_{00}^{(1)}$. The associated marginal distributions are}
\begin{subequations}
\label{marg_Sonine}
\beq
\frac{\phi_{c}(c)}{\phi_{c,M}(c)}=1+a_{20}^{(0)}\frac{{15}-20c^2+4c^4}{8},\quad
\frac{\phi_{w}(w)}{\phi_{w,M}(w)}=1+a_{02}^{(0)}\frac{{15}-20w^2+4w^4}{8},
\label{R_c}
\eeq
\bal
\frac{\phi_{cw}(c^2w^2)}{\phi_{cw,M}(c^2w^2)}=&1+
\frac{a_{20}^{(0)}+a_{02}^{(0)}}{8}\left[15+4c^2w^2-16cw\frac{K_1(2cw)}{K_0(2cw)}\right]\nn &
+\frac{a_{11}^{(0)}}{4}\left[9+4c^2w^2-12cw\frac{K_1(2cw)}{K_0(2cw)}\right],
\label{R_cw}
\eal
\bal
\frac{\phi_{\mathbf{c}\cdot\mathbf{w}}((\mathbf{c}\cdot\mathbf{w})^2)}{\phi_{\mathbf{c}\cdot\mathbf{w},M}((\mathbf{c}\cdot\mathbf{w})^2)}
=&1+
\frac{a_{20}^{(0)}+a_{02}^{(0)}}{8}\left[3+4(\mathbf{c}\cdot\mathbf{w})^2-
12|\mathbf{c}\cdot\mathbf{w}|\frac{K_0(2|\mathbf{c}\cdot\mathbf{w}|)}{K_1(2|\mathbf{c}\cdot\mathbf{w}|)}\right]\nn&
+\frac{a_{11}^{(0)}}{4}\left[1+4(\mathbf{c}\cdot\mathbf{w})^2-
8|\mathbf{c}\cdot\mathbf{w}|\frac{K_0(2|\mathbf{c}\cdot\mathbf{w}|)}{K_1(2|\mathbf{c}\cdot\mathbf{w}|)}\right]
\nn&
-\frac{a_{00}^{(1)}}{2}\left[1-2(\mathbf{c}\cdot\mathbf{w})^2+
|\mathbf{c}\cdot\mathbf{w}|\frac{K_0(2|\mathbf{c}\cdot\mathbf{w}|)}{K_1(2|\mathbf{c}\cdot\mathbf{w}|)}\right],
\label{R_cdw}
\eal
\beq
\frac{\phi_\vartheta(\cos^2\vartheta)}{\phi_{\vartheta,M}(\cos^2\vartheta)}=
1+\frac{9a_{00}^{(1)}}{8}\left(3\cos^2\vartheta-1\right).
\label{R_theta}
\eeq
\end{subequations}
This yields $\langle c^4\rangle=\frac{15}{4}(1+a_{20}^{(0)})$, $\langle w^4\rangle=\frac{15}{4}(1+a_{02}^{(0)})$, $\langle c^2w^2\rangle=\frac{9}{4}(1+a_{11}^{(0)})$, $\langle (\mathbf{c}\cdot\mathbf{w})^2\rangle=\frac{3}{4}(1+a_{11}^{(0)}+\frac{5}{2}a_{00}^{(1)})$, and $\langle\cos^2\vartheta\rangle=\frac{1}{3}(1+\frac{9}{10}a_{00}^{(1)})$, as expected. The latter equality implies $b=a_{00}^{(1)}$, in consistency with the neglect of $a_{jk}^{(\ell)}$ with $j+k+2\ell\geq 3$ in Eq.\ \eqref{5.13}. If, additionally, terms nonlinear in the cumulants are neglected in Eq.\ \eqref{5.14}, we have
\beq
\label{hba}
{h\simeq b\simeq a_{00}^{(1)}}
\eeq
{in the Sonine approximation \eqref{phi2}.
Thus, if $a_{00}^{(1)}<0$ (as will be seen to be the case for most values of $\alpha$ and $\beta$ in the case of uniform spheres), we can expect that $\langle\cos^2\vartheta\rangle<\frac{1}{3}$. This implies a tendency of the vectors $\mathbf{c}$ and $\mathbf{w}$ to favor quasinormal mutual orientations (``lifted-tennis-ball'' effect).}

If Eq.\ \eqref{phi2} is inserted into Eq.\ \eqref{mupq} and terms nonlinear in $a_{jk}^{(\ell)}$ are neglected, one  obtains explicit (yet not exact) expressions of $\mu_{pq}^{(r)}$ with $p+q+2r=2$ and $ 4$. Those expressions were first displayed  in the Supplemental material of Ref.\ \onlinecite{VSK14} but, for the sake of completion, they can be found here in the Appendix.
Once Eqs.\ \eqref{22}--\eqref{30n} are introduced into Eqs.\  \eqref{thetadt} and \eqref{evocum}, the resulting closed set of six equations can be numerically solved for arbitrary initial values to get the time-dependent functions $\theta(\tau)$, $\gamma(\tau)$, $a_{20}^{(0)}(\tau)$, $a_{02}^{(0)}(\tau)$, $a_{11}^{(0)}(\tau)$, and $a_{00}^{(1)}(\tau)$.\cite{note_15_08_1} We will include in Sec.\ \ref{sec4}  graphs of those  functions and compare them with the ``exact'' numerical solution of the Boltzmann equation obtained by means of the DSMC method.\cite{B94}

As for the steady state, by applying the state-state conditions \eqref{cumss} and the Sonine expressions \eqref{22}--\eqref{30n}, we are able to obtain fully analytical explicit expressions for the temperature ratio $\theta$, the dimensionless noise intensity $\gamma$, and the four cumulants $a_{20}^{(0)}$, $a_{02}^{(0)}$, $a_{11}^{(0)}$, and $a_{00}^{(1)}$.
{The method is quite simple.\cite{note_15_08_1} First, the relations in Eq.\ \eqref{5.12}  are used to express the four cumulants in terms of the yet unknown quantity $\theta$. Inserting those expressions into $\mu_{02}^{(0)}=0$, one gets a closed quartic equation for $\theta$, whose physical solution is identified as the one closer to $\theta_M$ [see Eq.\ \eqref{thetaMax}]. Finally, $\gamma$ is obtained from $\gamma=\mu_{20}^{(0)}$.}

\begin{figure}[ht]
\includegraphics[height=\width]{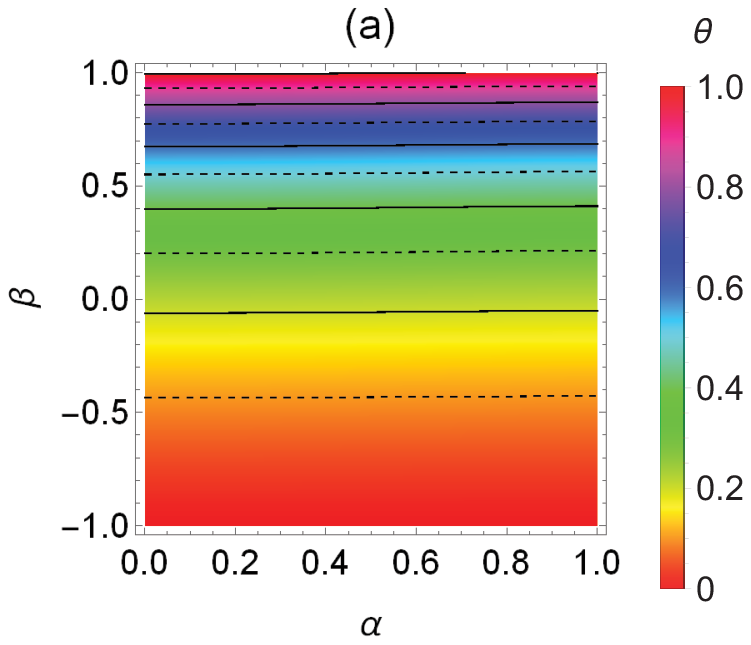}\hspace{1cm}
\includegraphics[height=\width]{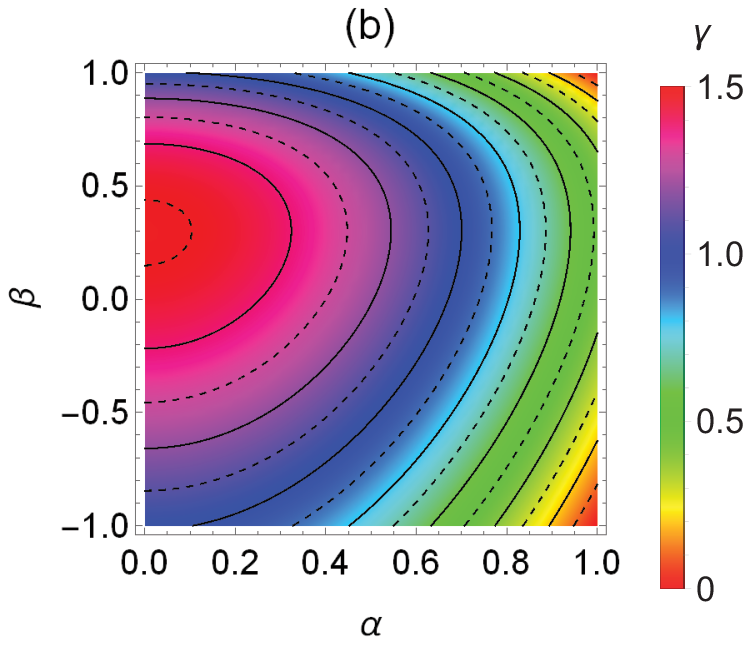}\\
\vspace{0.2cm}
\includegraphics[height=\width]{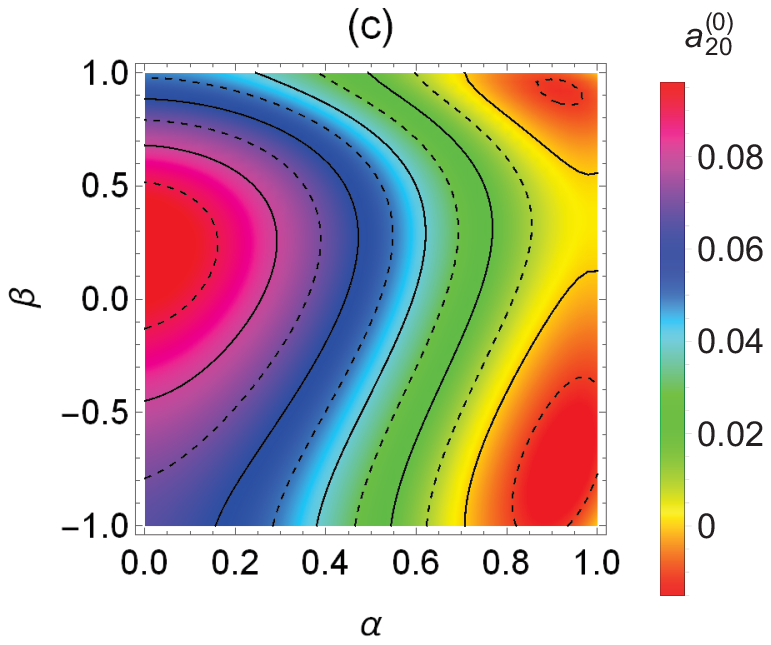}\hspace{1cm}
\includegraphics[height=\width]{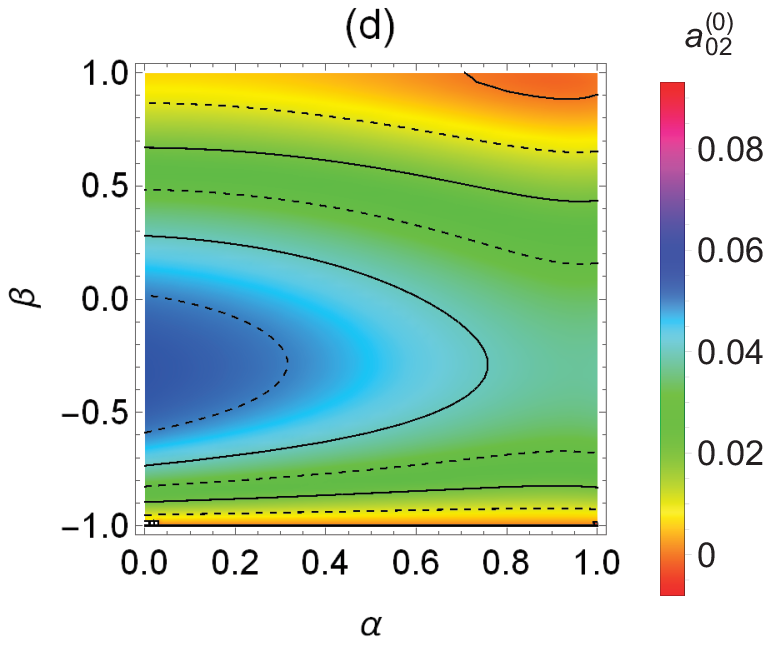}\\
\vspace{0.2cm}
\includegraphics[height=\width]{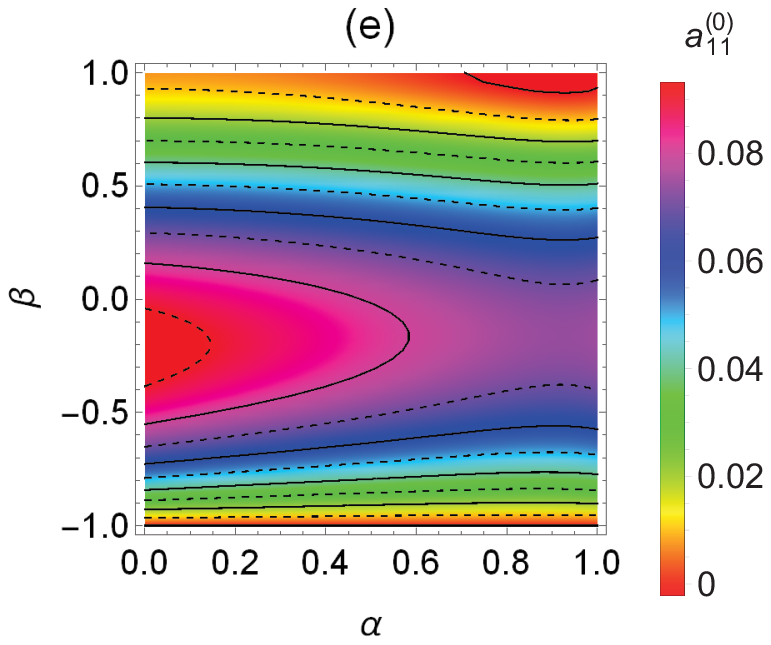}\hspace{1cm}
\includegraphics[height=\width]{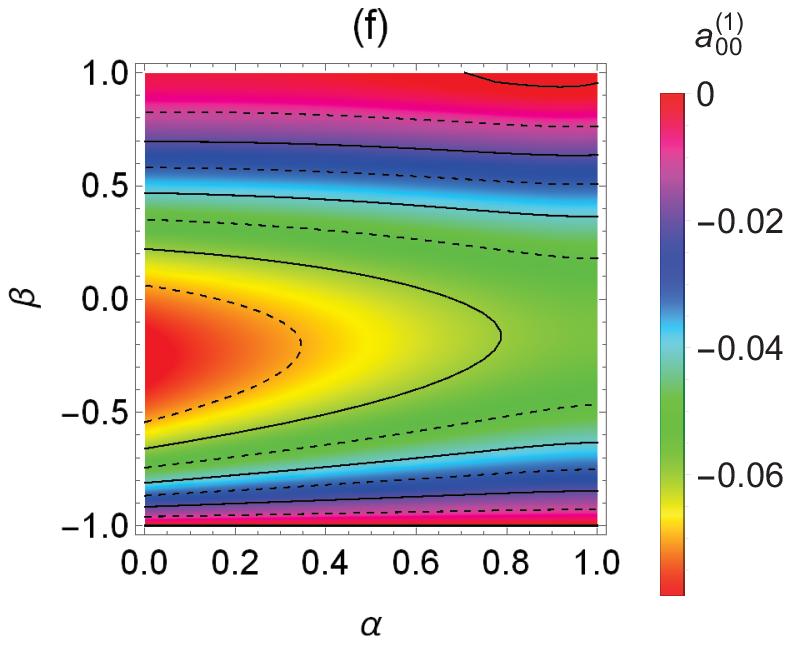}
\caption{{Density plots}, for uniform spheres ($\kappa=\frac{2}{5}$), of the steady-state values of  (a) the temperature ratio $\theta=T_r/T_t$, (b) the dimensionless noise intensity $\gamma$ [see Eq.\ \protect\eqref{Gamma}],  (c) the {translational kurtosis} $a_{20}^{(0)}$ [see Eq.\ \protect\eqref{a20}], (d) the {rotational kurtosis} $a_{02}^{(0)}$ [see Eq.\ \protect\eqref{a20}], (e) the {scalar correlation} cumulant $a_{11}^{(0)}$ [see Eq.\ \protect\eqref{b}], and (f) the {orientational correlation} cumulant $a_{00}^{(1)}$ [see Eq.\ \protect\eqref{b}]. {The contour lines correspond to (a) $\theta=0.1,0.2,\ldots,0.9$, (b) $\gamma=0.1,0.2,\ldots,1.5$, (c) $a_{20}^{(0)}=-0.01,0.00,\ldots,0.09$, (d) $a_{02}^{(0)}=0.00,0.01,\ldots,0.05$,  (e) $a_{11}^{(0)}=0.00,0.01,\ldots,0.09$, and (f) $a_{00}^{(1)}=-0.07,-0.06,\ldots,0.00$.}
\label{maps}}
\end{figure}

Figure \ref{maps} presents {density plots} of the steady-state values for uniform spheres ($\kappa=\frac{2}{5}$) of $\theta$, $\gamma$, $a_{20}^{(0)}$, $a_{02}^{(0)}$, $a_{11}^{(0)}$, and $a_{00}^{(1)}$.
{In Fig.\ \ref{maps}(a) we may point out that iso-$\theta$ curves are almost perpendicular to the $\beta$ axis, indicating that $\theta$ depends \emph{almost} entirely on the coefficient of tangential restitution and hardly on $\alpha$, as anticipated by the Maxwellian approximation  $\theta_M$ in Eq.\ \eqref{thetaMax}. Also, $\theta<1$, except, and this is not visible in Fig.\ \ref{maps}(a), if $\beta>0.994\,336$ and $0\leq\alpha<\alpha_1(\beta)$, where $\alpha_1(0.994\,336)=0$ and $\alpha_1(1)=1/\sqrt{2}$. The maximum value $\theta_{\max}=1.009\,85$ takes place at $\alpha=0$ and $\beta=1$, i.e., in the completely inelastic and rough limit. The behavior of $\gamma$ observed in Fig.\ \ref{maps}(b) is in good agreement with the one expected from the Maxwellian approximation $\gamma_M$ in Eq.\ \eqref{thetaMax}. The latter predicts, at fixed $\alpha$, a maximum value  $\gamma_{M,\max}\simeq 1.4853 -\alpha^2$ at $\beta\simeq 0.3033$. In the Sonine approximation, we find that the location and value of the maximum slightly changes from $\beta\simeq 0.301$ and $\gamma_{\max}+\alpha^2\simeq 1.512$ at $\alpha=0$ to
$\beta\simeq 0.304$ and $\gamma_{\max}+\alpha^2\simeq 1.486$ at $\alpha=1$.}

{Since $\theta$ and $\gamma$ are well represented by the Maxwellian approximation \eqref{thetaMax}, the most interesting panels in Fig.\ \ref{maps} are those related to the cumulants. The first feature to be noted is that all of them have a relatively small magnitude $|a_{jk}^{(\ell)}|<0.1$, implying that a good performance of the Sonine approximation \eqref{phi2}, and hence the validity of \eqref{hba}, can be expected. This expectation will be confirmed in Sec.\ \ref{sec4}.}
We can also see that, at given $\alpha$, the maximum value of $|a_{jk}^{(\ell)}|$ is typically reached in the region of intermediate roughness ($\beta\sim0$), which differs from the behavior detected for the HCS, where the  maximum values (which are also much larger than here) occur in the region close to the smooth limit.\cite{VSK14}

{Figure \ref{maps}(c) shows that the translational kurtosis $a_{20}^{(0)}$ takes negative values in the top and bottom right corners of the map. In contrast, the  region where the rotational kurtosis $a_{02}^{(0)}$ is negative reduces to a small top-right corner ($\alpha,\beta\lesssim 1$) [see Fig.\ \ref{maps}(d)]. The analogous top-right corners where $a_{11}^{(0)}<0$ and $a_{00}^{(1)}>0$, respectively, are even smaller in the cases of the correlation cumulants  $a_{11}^{(0)}$ and, especially,  $a_{00}^{(1)}$ [see Figs.\ \ref{maps}(e) and \ref{maps}(f)].}

{Therefore, the typical non-Maxwellian features of the joint velocity distribution function $\phi(\mathbf{c},\mathbf{w})$ that can be extracted from Figs.\ \ref{maps}(c)--\ref{maps}(f) are: (i) platykurtic translational ($a_{20}^{(0)}>0\Rightarrow \langle c^4\rangle>\frac{15}{4}$) and rotational ($a_{02}^{(0)}>0\Rightarrow \langle w^4\rangle>\frac{15}{4}$) velocity distributions; (ii) particles with larger (smaller) translational velocities tend to have larger (smaller) rotational velocities ($a_{11}^{(0)}>0\Rightarrow \langle c^2 w^2\rangle>\langle c^2\rangle\langle w^2\rangle=\frac{9}{4}$); and (iii) quasinormal orientations  between the vectors $\mathbf{c}$ and $\mathbf{w}$  tend to be favored against quasiparallel orientations ($a_{00}^{(1)}<0\Rightarrow \langle \cos^2\vartheta\rangle<\frac{1}{3}$).
Exceptions to these general features in the cases of $a_{02}^{(0)}$, $a_{11}^{(0)}$, and $a_{00}^{(1)}$ are limited to small regions with $\alpha\lesssim 1$ and $\beta\lesssim 1$ (top-right corners). For the kurtosis $a_{20}^{(0)}$, the size of the top-right exception region is significantly larger and, moreover, another even larger bottom-right corner appears.
}

{It is worth remarking that, while there still exists  a prevalence of $\langle \cos^2\vartheta\rangle<\frac{1}{3}$  (lifted-tennis-ball effect)  for the unforced granular gas in the HCS,\cite{BPKZ07,KBPZ09,RA14,VSK14}  a significantly larger  region with $\langle \cos^2\vartheta\rangle>\frac{1}{3}$ (``cannon-ball'' effect) appears close to $\alpha,\beta\lesssim 1$ and a second  region with the same behavior appears in the opposite perfectly smooth limit ($\alpha\lesssim 1$, $\beta\gtrsim -1$).
The counterpart of this second region is completely absent in the case of a heated granular gas considered here. This is closely related  to the fact that, while the quasismooth limit $\beta\to -1$ is singular in the HCS,\cite{BPKZ07,VSK14} it becomes regular in the heated case. Taking carefully the limit $\beta\to -1$ in the analytical expressions for $\theta$, $\gamma$, and $a_{jk}^{(\ell)}$, one gets\cite{SKS11}}
\begin{subequations}
\beq
{\theta\to\frac{61-45\al+ 6\al^2(1-\al)}{15 - 11 \al +2 \al^2(1-\al)}\frac{1+\beta}{8}{\frac{\kappa}{\kappa+1}}\to 0,\quad
\gamma\to4(1-\al^2)\frac{61 - 45\al+6\al^2(1-\al)}{241 - 177 \al +30 \al^2(1-\al)}},
\label{5.20}
\eeq
\beq
{a_{20}^{(0)}\to16(1-\al)\frac{1 - 2\al^2}{241 - 177 \al +30 \al^2(1-\al)}, \quad a_{02}^{(0)}\to 0,\quad
a_{11}^{(0)}\to 0,\quad a_{00}^{(1)}\to 0}.
\label{5.19}
\eeq
\end{subequations}
{The expression of $a_{20}^{(0)}$ in Eq.\ \eqref{5.19} coincides with the one derived directly for strict smooth spheres ($\beta=-1$).\cite{vNE98} Interestingly, in the latter case the rotational distribution
function (and hence the temperature $\Tr$) is not uniquely defined since it preserves with time its initial arbitrary form. On the other hand, the limit $\beta\to -1$ shows that the the rotational degrees of freedom become decoupled from the translational ones ($a_{11}^{(0)}\to 0$, $a_{00}^{(1)}\to 0$) but the rotational distribution function tends to a Maxwellian $a_{02}^{(0)}\to 0$ with a temperature much smaller than the translational one ($\Tr/\Tt\to 0$).}

Of course the theoretical results presented in this section are valid only if they agree with a direct comparison with a solution of the Boltzmann equation \eqref{BE} {free from any additional hypotheses}. This comparison will be done in Sec.\ \ref{sec4}.

\section{Comparison of  theoretical results with DSMC data}
\label{sec4}

We display in this section detailed comparisons of the theoretical results with  accurate DSMC data, as obtained from a code we wrote specifically for the Boltzmann equation \eqref{BE} for this system. For data noise reduction, we have used $2\times10^6$ simulation particles for all cases in this work. More details on the implementation of DSMC in a heated granular gas can be found in Ref.\ \onlinecite{MS00}.
We fixed the value $\kappa=\frac{2}{5}$ (uniform spheres) for the reduced moment of inertia and varied either the coefficient of normal restitution $\alpha$ (at fixed  $\beta$) or the coefficient of tangential restitution $\beta$ (at fixed $\alpha$). {The noise intensity parameter  was typically chosen  as $\chi_0^2=\frac{24}{5}n\pi\sigma^2[\Tt(0)/m]^{3/2}$, which corresponds to $\gamma(0)=\frac{18}{5}\sqrt{\pi}\simeq 6.381$, but other values were also considered to check the independence of the steady state on the precise value of $\chi_0^2$.} In all cases, the particle velocities were initially sampled from a Maxwellian distribution [i.e., $a_{jk}^{(\ell)}(0)=0$] with a common temperature $\Tr(0)=\Tt(0)$ [i.e., $\theta(0)=1$]. For each different system, the corresponding data for both transient and steady states were recorded.

\subsection{Transient behavior}

\begin{figure}
\includegraphics[height=\widthone]{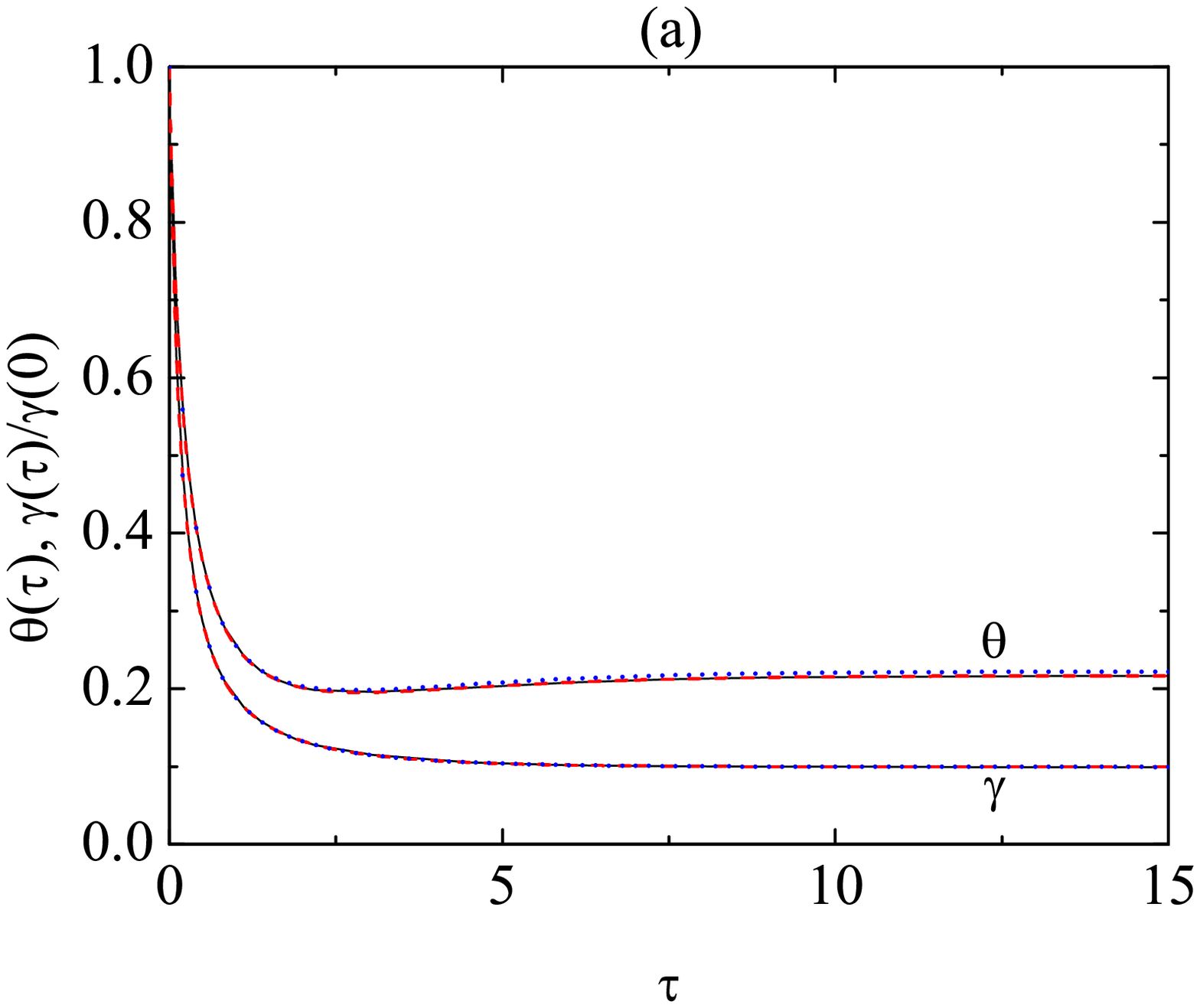}
\includegraphics[height=\widthone]{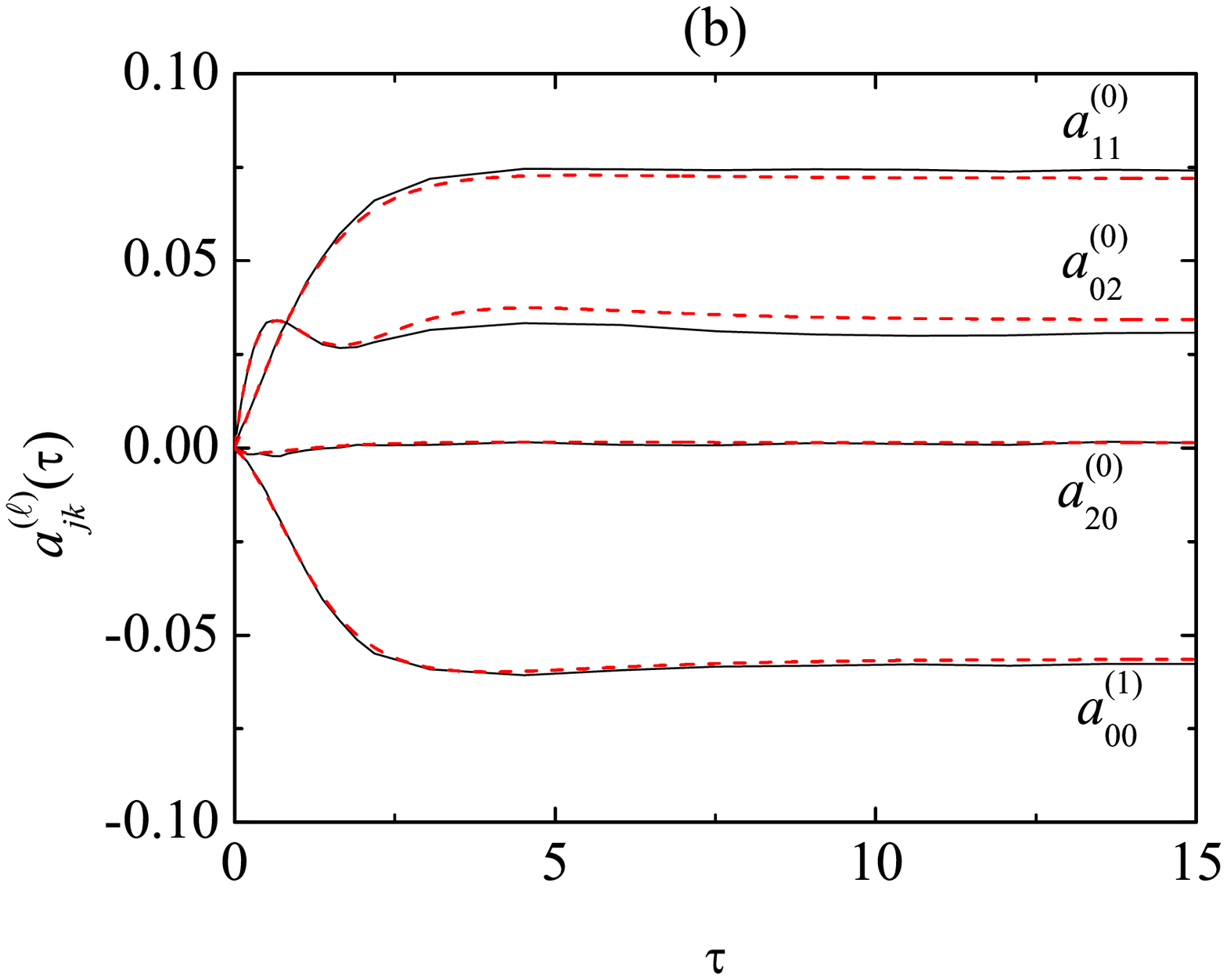}
\caption{Evolution, for  $\alpha=0.9$, $\beta=0$, and {$\gamma(0)=\frac{18}{5}\sqrt{\pi}\simeq 6.381$} of (a) temperature ratio $\theta(\tau)$ and reduced noise intensity (normalized to its initial value) $\gamma(\tau)/\gamma(0)$, and (b)  cumulants $a_{jk}^{(\ell)}(\tau)$. Time is measured in units of number of collisions per particle, $\tau$. Solid lines stand for DSMC results, {dashed lines for the Sonine approximation, and dotted lines [only in panel (a)] for the Maxwellian approximation}.}
\label{evol}
\end{figure}

As an illustration of the relaxation towards the steady state, and for the case $\alpha=0.9$, $\beta=0$, and {$\gamma(0)=\frac{18}{5}\sqrt{\pi}\simeq 6.381$}, we show in Fig.\ \ref{evol}  the evolution of the  temperature ratio  $\theta(\tau)$ and the dimensionless noise intensity $\gamma(\tau)$ in panel (a), and of the four cumulants $a_{20}^{(0)}(\tau)$, $a_{02}^{(0)}(\tau)$, $a_{11}^{(0)}(\tau)$, and $a_{00}^{(1)}(\tau)$  in panel (b).
{A very good agreement between theory and simulation is observed. In the case of $\gamma(\tau)$, the Maxwellian and Sonine approximations give results practically indistinguishable from the DSMC ones. For $\theta(\tau)$, the Sonine approximation is  still virtually perfect, while the Maxwellian  one tends to slightly overestimate  this quantity. The initial reduced white noise intensity $\gamma(0)=\frac{18}{5}\sqrt{\pi}$ is larger than the initial cooling rate $\mu_{20}^{(0)}(0)$, and therefore the external-force heating effect dominates over the inelastic cooling. As a consequence, $\gamma(\tau)$ monotonically decays in time [see Eq.\ \eqref{thetadt}] until the steady state is reached (after about $10$ collisions per particle). In this respect, note that, according to Eq.\ \eqref{Gamma}, the normalized quantity $\gamma(\tau)/\gamma(0)$ plotted in Fig.\ \ref{evol}(a) is equivalent to $[\Tt(\tau)/\Tt(0)]^{-3/2}$. Thus, the steady-state value $\gamma(\tau)/\gamma(0)\to 0.099$ implies $\Tt(\tau)/\Tt(0)\to 4.66$. The heating is much less efficient in the case of $\Tr$ since the stochastic force acts directly on the translational velocity only. Therefore, the rotational-to-translational temperature ratio decays (but non-monotonically) to $\Tr(\tau)/\Tt(\tau)\to 0.217$ and the rotational temperature has hardly increased to $\Tr(\tau)/\Tr(0)\to 1.010$.}

{The evolution of the cumulants is much less intuitive. As can be observed in Fig.\ \ref{evol}(b), for the case under consideration ($\alpha=0.9$, $\beta=0$) the translational kurtosis $a_{20}^{(0)}$ hardly increases, the rotational kurtosis $a_{02}^{(0)}$ grows non-monotonically, and the more relevant non-Maxwellian properties are associated with the correlation quantities $a_{11}^{(0)}$ and $a_{00}^{(1)}$. All these features are well captured by our Sonine approximation, although it tends to slightly overestimate $a_{02}^{(0)}$ and underestimate $a_{11}^{(0)}$ and  $|a_{00}^{(1)}|$.}

\subsection{Steady state}
\begin{figure}
\includegraphics[height=\widthtwocm]{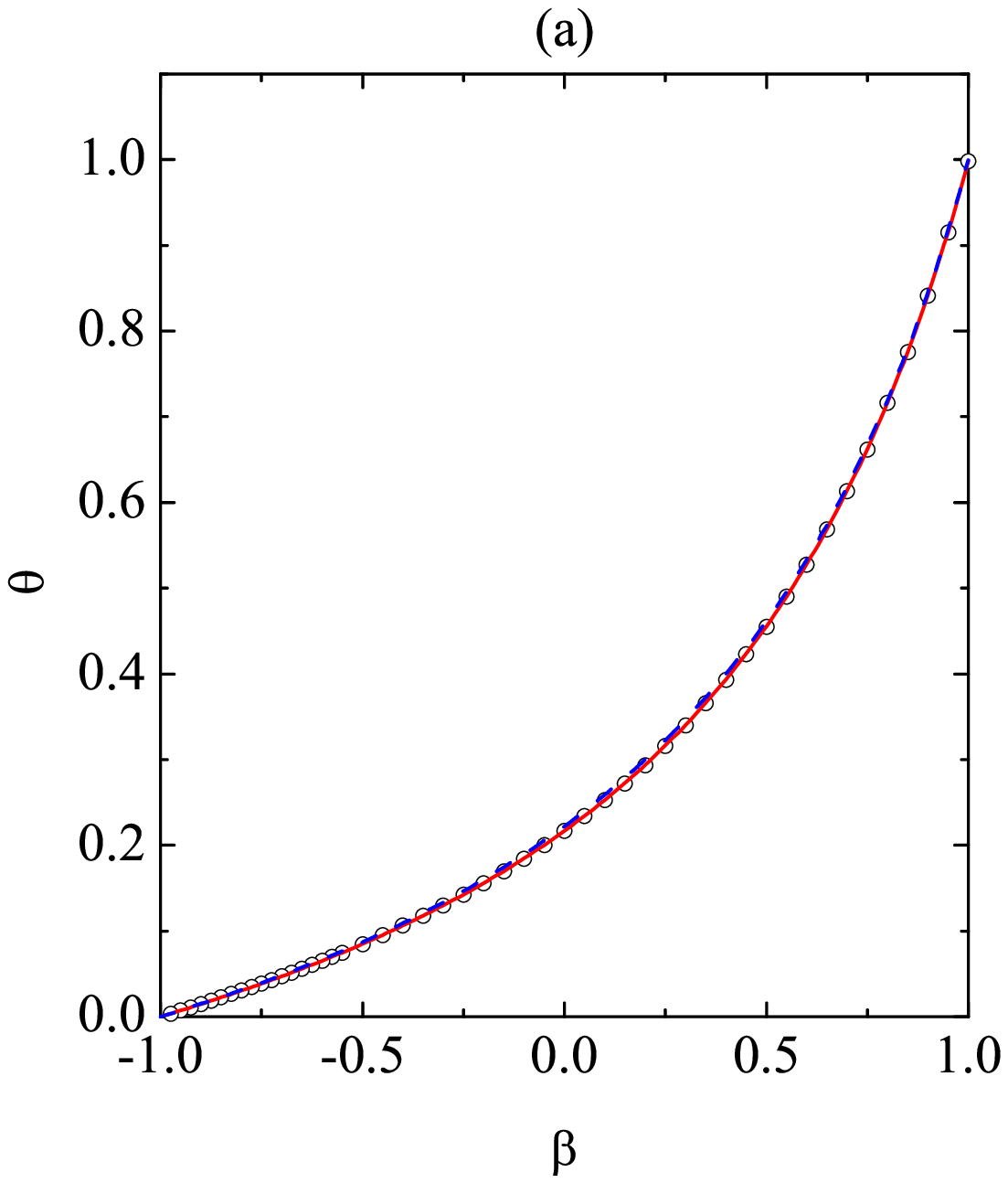}
\includegraphics[height=\widthtwocm]{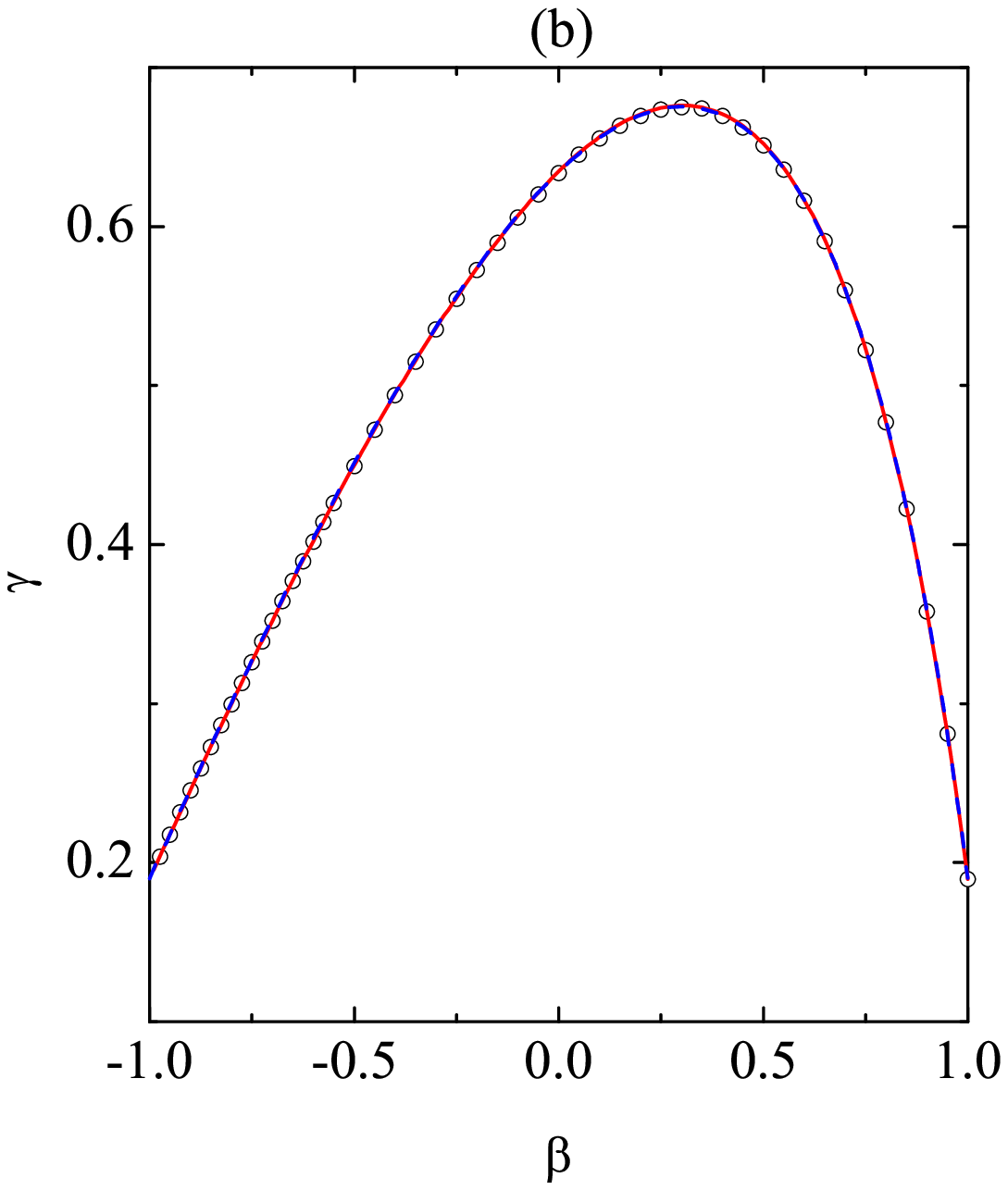}
\includegraphics[height=\widthtwocm]{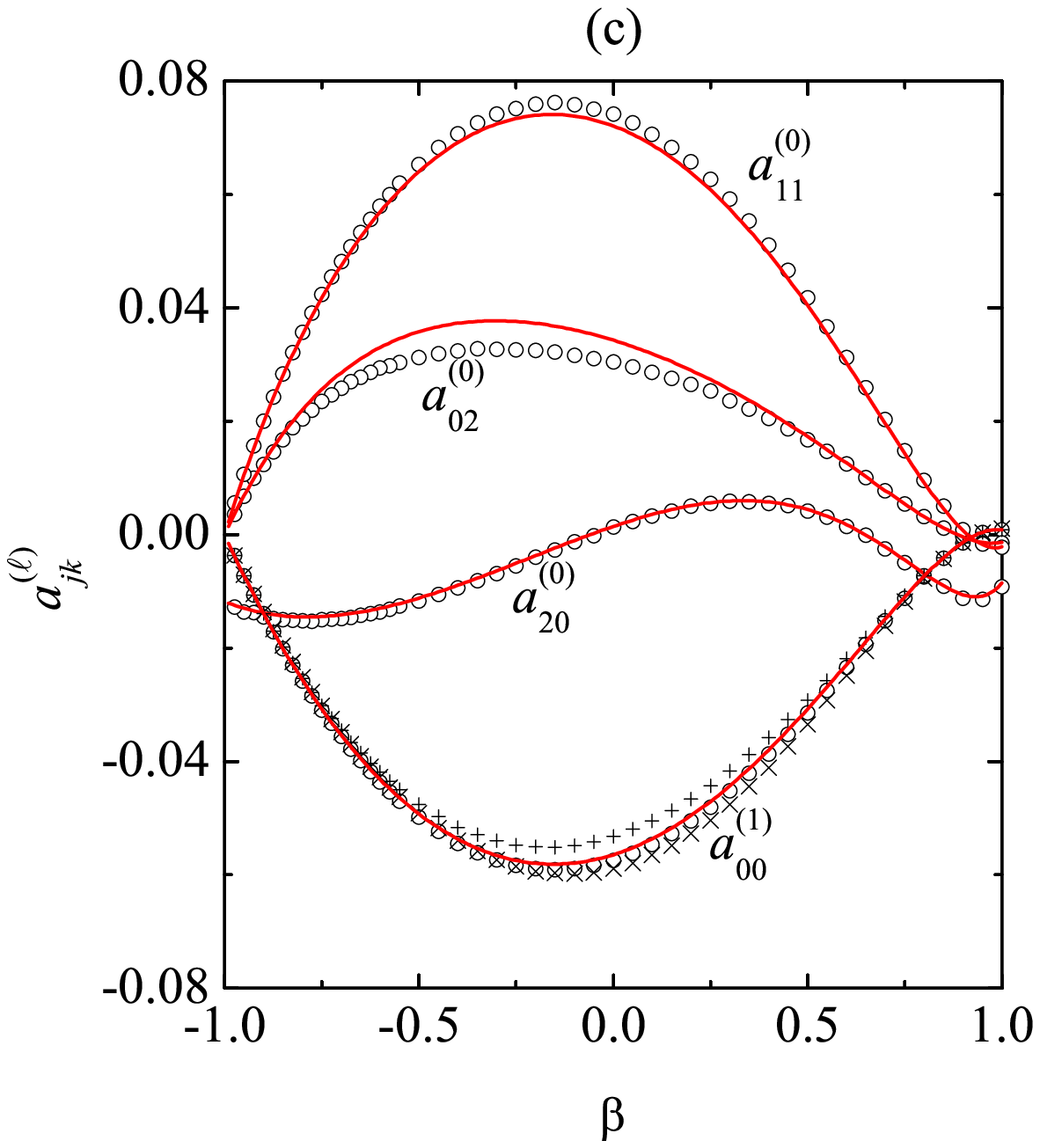}
\caption{{Steady-state values of (a) the temperature ratio $\theta$, (b) the reduced noise intensity $\gamma$, and (c) the cumulants $a_{jk}^{(\ell)}$ as functions of $\beta$ for $\alpha=0.9$. Circles stand for DSMC data [with $\gamma(0)=\frac{18}{5}\sqrt{\pi}\simeq 6.381$], solid lines for the Sonine approximation, and dashed lines [only in panels (a) and (b)] for the Maxwellian approximation. The DSMC data for the quantities $b$ ($\times$) and $h$ ($+$) [see Eq.\ \eqref{eqb}] are also included in panel (c).}}
\label{alfa09}
\end{figure}

{Now we focus on the steady state. The influence of the coefficient of tangential restitution $\beta$ on $\theta$, $\gamma$, and $a_{jk}^{(\ell)}$ at a fixed coefficient of normal restitution $\alpha=0.9$ is shown in Fig.\ \ref{alfa09}. The DSMC data for the correlation quantities $b$ and $h$ [see Eq.\ \eqref{eqb}] are also plotted in panel (c). Panels (a) and (b) confirm the excellent performance of the simple Maxwellian approximation \eqref{thetaMax} and of the more elaborate Sonine approximation in what concerns $\theta$ and $\gamma$. The temperature ratio $\theta$ monotonically increases from $\theta\to 0$ in the smooth limit $\beta\to -1$ [see \eqref{5.20}] to $\theta\simeq 1$ at the perfectly rough case $\beta=1$. Although not visible on the scale of Fig.\ \ref{alfa09}(a), the DSMC value at $\beta=1$ is actually $\theta=0.998$.
The quantity $\gamma$ exhibits a non-monotonic dependence on $\beta$, as already discussed in connection with Fig.\ \ref{maps}(b), with a maximum $\gamma_{\max}\simeq 0.675$ at $\beta\simeq 0.3$.}

{Regarding the cumulants, the behaviors observed in Fig.\ \ref{alfa09}(c) agree with what might have been anticipated by imagining a vertical slice $\alpha=0.9$ in Figs.\ \ref{maps}(c)--(f). While $a_{02}^{(0)}$, $a_{11}^{(0)}$, and $|a_{00}^{(1)}|$ vanish at $\beta\to -1$, are very small at $\beta=1$, and present maxima in the medium-roughness region $\beta\sim 0$, the translational kurtosis $a_{20}^{(0)}$ has a more complex behavior with a (positive) maximum at $\beta\simeq 0.34$ and two (negative) local minima at $\beta\simeq -0.79$ and $\beta\simeq 0.93$. We can also observe that the expectations \eqref{hba} are indeed satisfied. Actually, the only visible differences between $a_{00}^{(1)}$, $b$, and $h$ occur in the region $\beta\sim 0$, where one has $|h|\lesssim |a_{00}^{(1)}|\lesssim |b|$. As is apparent from Fig.\ \ref{alfa09}(c), the performance of the Sonine approximation is very good, although small discrepancies, especially in the case of the rotational kurtosis $a_{02}^{(0)}$, appear in the region of medium roughness ($\beta\sim 0$), where the magnitudes of the cumulants are higher. Although hardly visible in Fig.\ \ref{alfa09}(c), the theory succeeds in predicting that $a_{02}^{(1)}<0$ (leptokurtic rotational velocity distribution), $a_{11}^{(0)}<0$ (high $c$ correlated with low $w$, and viceversa), and $a_{00}^{(1)}>0$ (cannon-ball effect) in the regions $0.89\lesssim \beta\leq 1$, $0.92\lesssim \beta\leq 1$, and $0.94\lesssim \beta\leq 1$, respectively. Partially due to the fact that the translational kurtosis $a_{20}^{(0)}$ is the cumulant of smaller magnitude in the central region, its Sonine prediction is excellent for all $\beta$.}

\begin{figure}
\includegraphics[height=\widthtwocm]{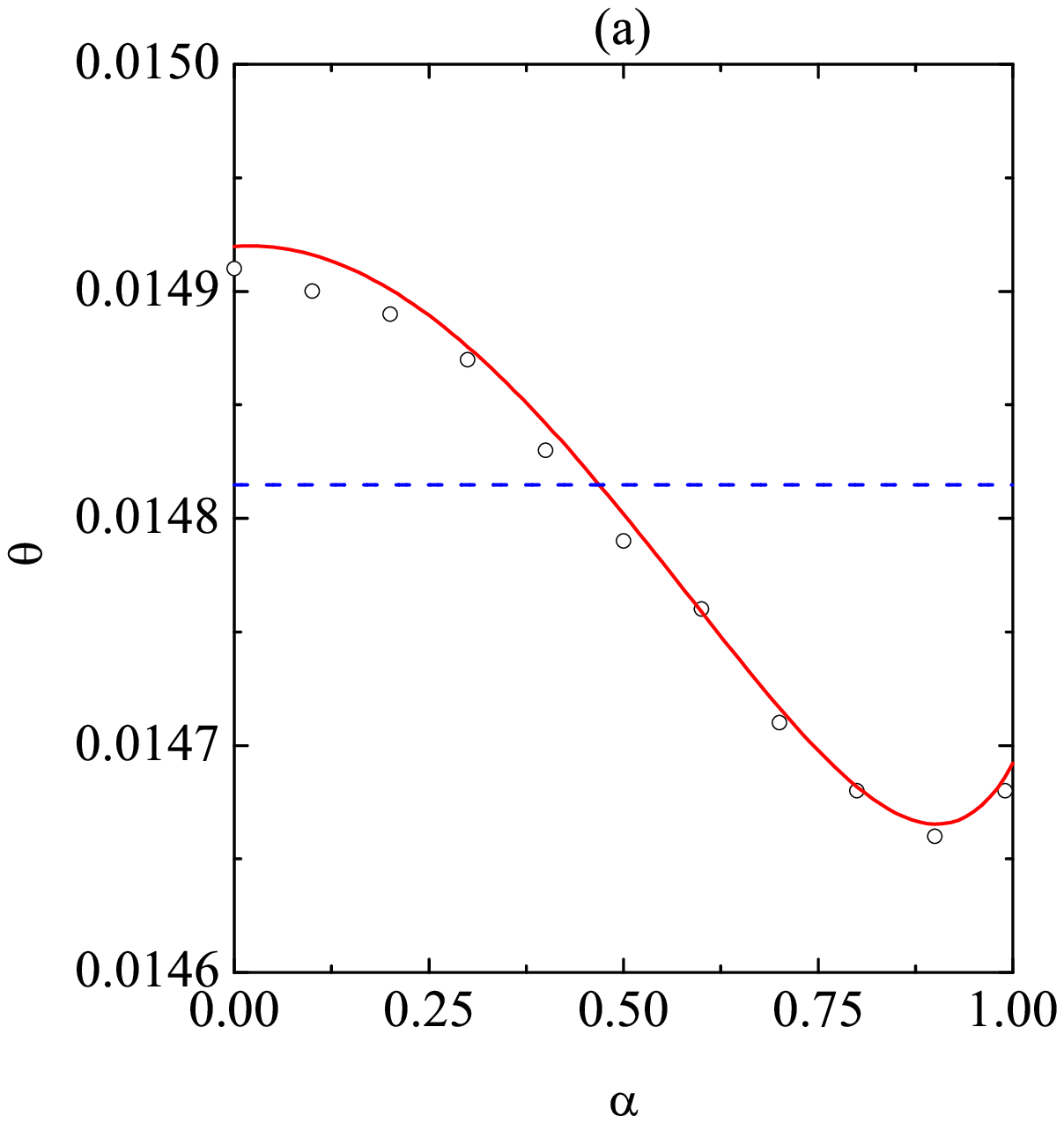}
\includegraphics[height=\widthtwocm]{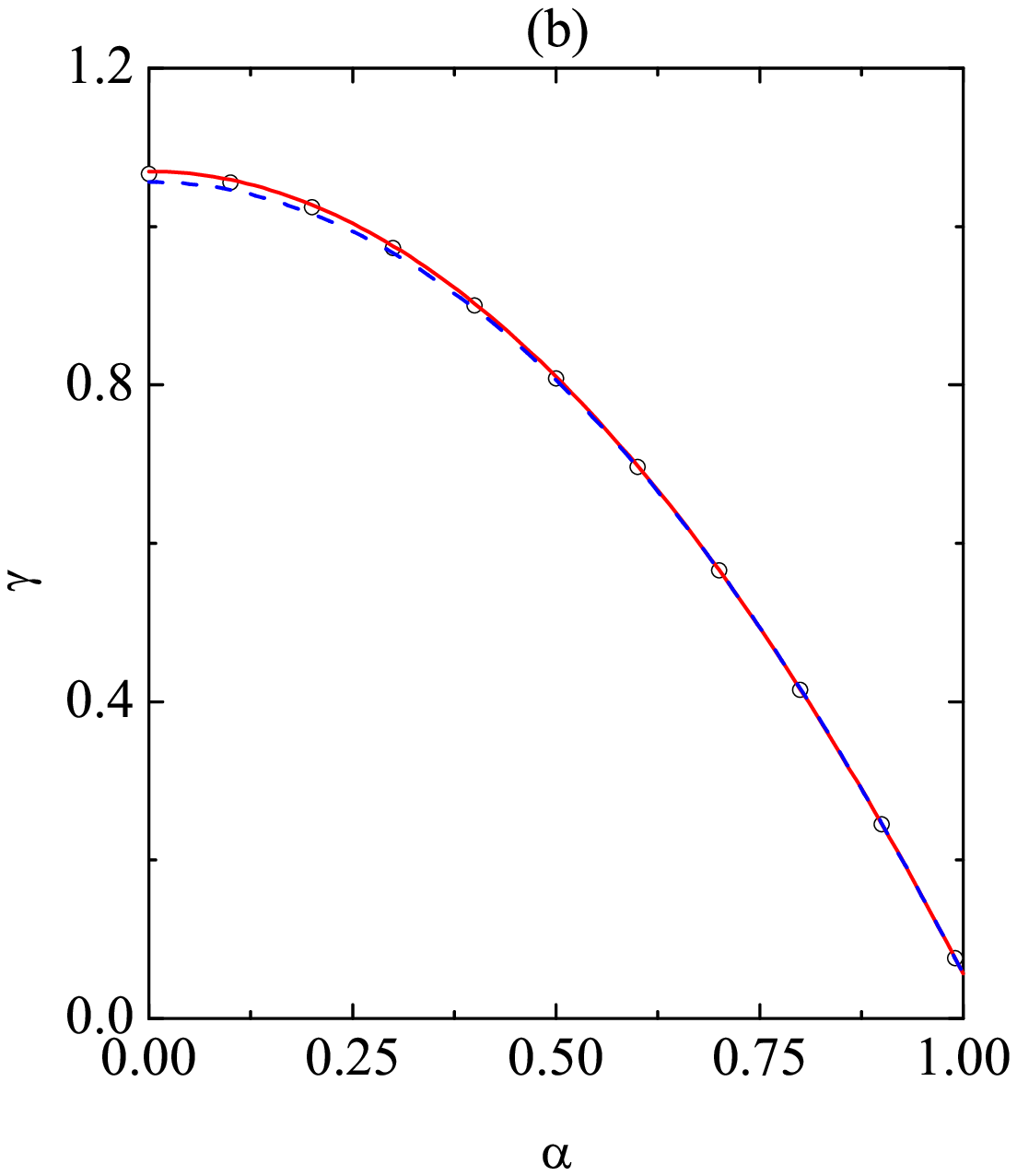}
\includegraphics[height=\widthtwocm]{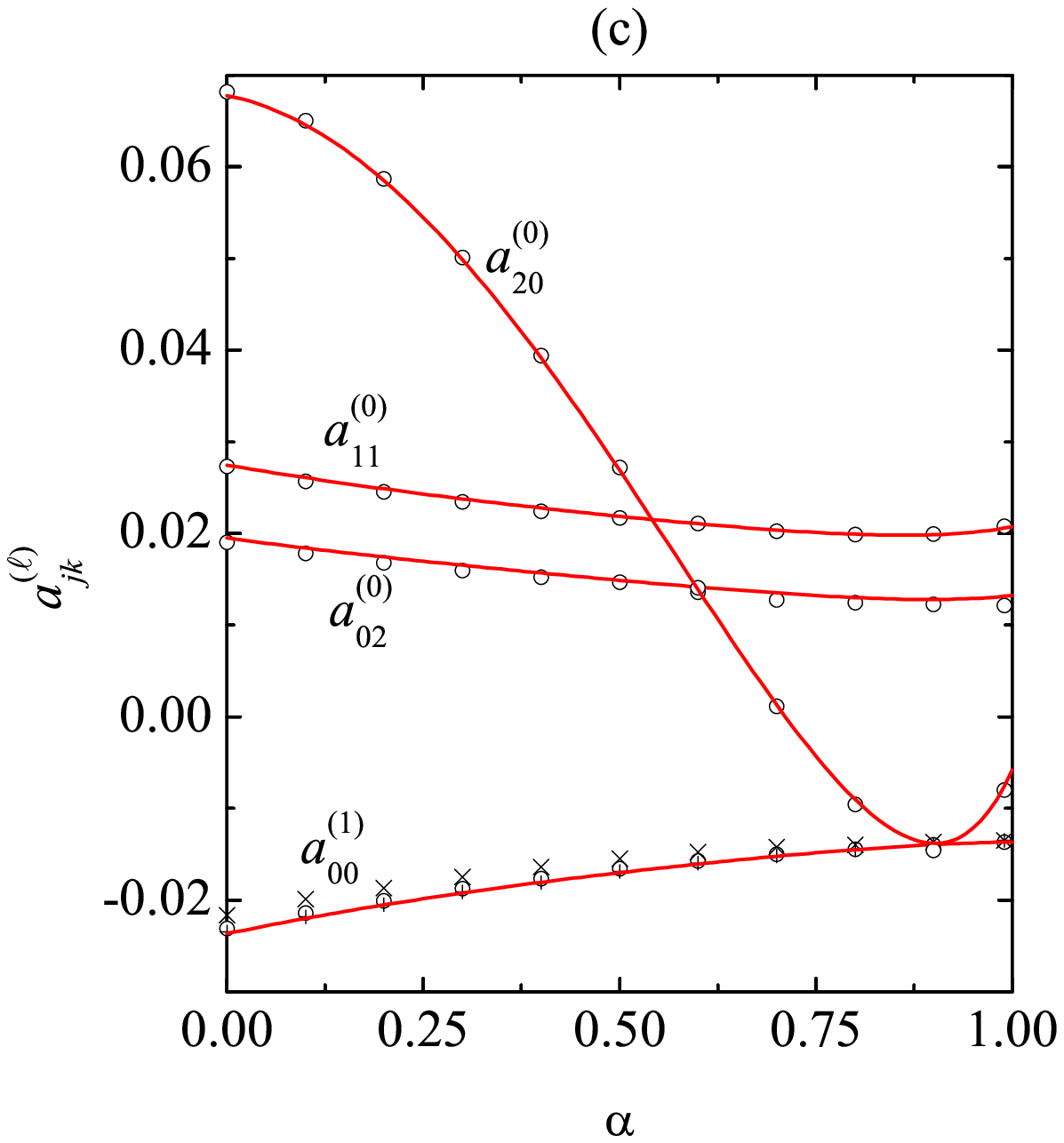}
\caption{
{Steady-state values of (a) the temperature ratio $\theta$, (b) the reduced noise intensity $\gamma$, and (c) the cumulants $a_{jk}^{(\ell)}$ as functions of $\alpha$ for $\beta=-0.9$. Circles stand for DSMC data [with $\gamma(0)=\frac{18}{5}\sqrt{\pi}\simeq 6.381$], solid lines for the Sonine approximation, and dashed lines [only in panels (a) and (b)] for the Maxwellian approximation. The DSMC data for the quantities $b$ ($\times$) and $h$ ($+$) [see Eq.\ \eqref{eqb}] are also included in panel (c).}
\label{betam09}}
\end{figure}
\begin{figure}
\includegraphics[height=\widthtwocm]{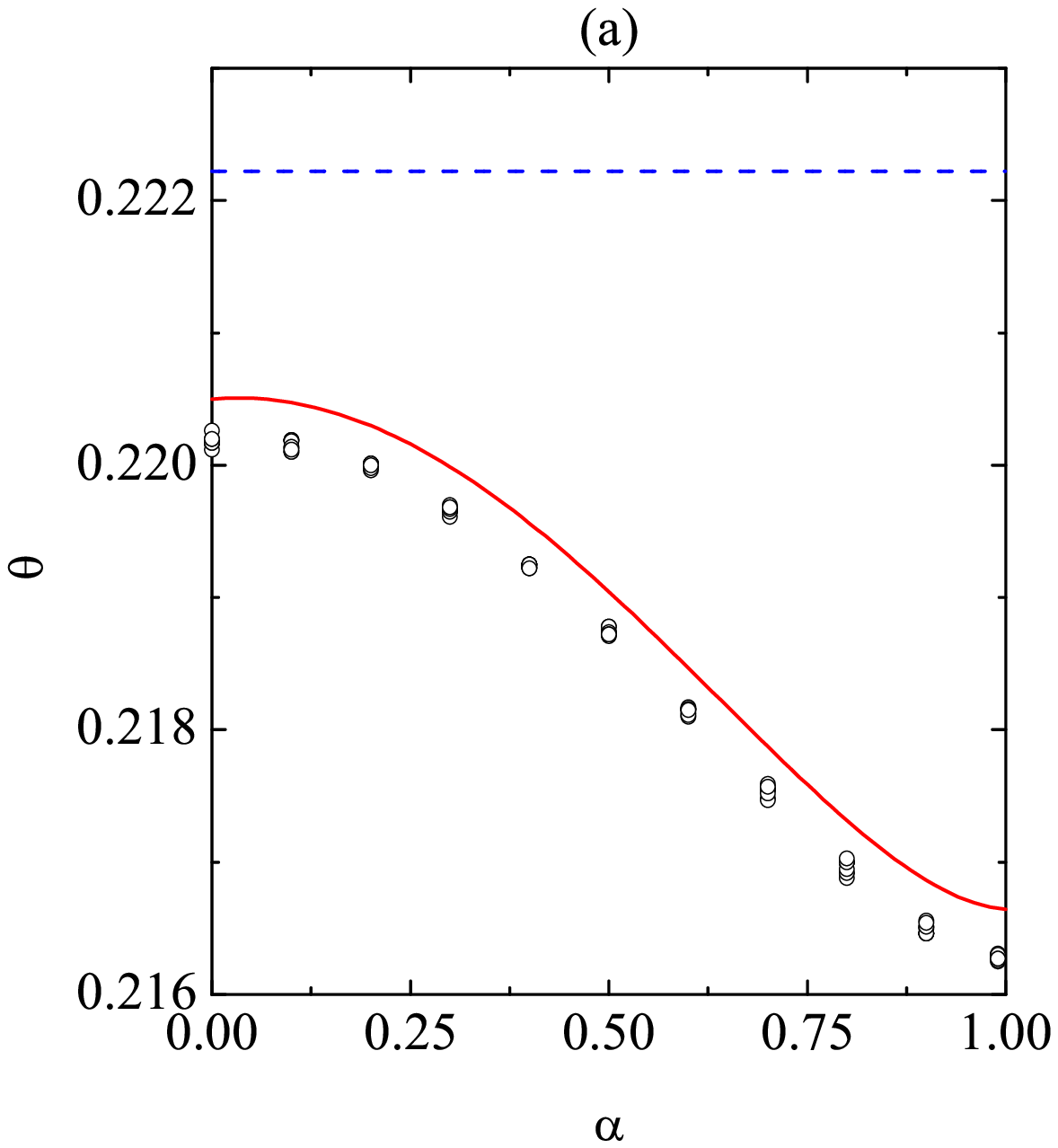}
\includegraphics[height=\widthtwocm]{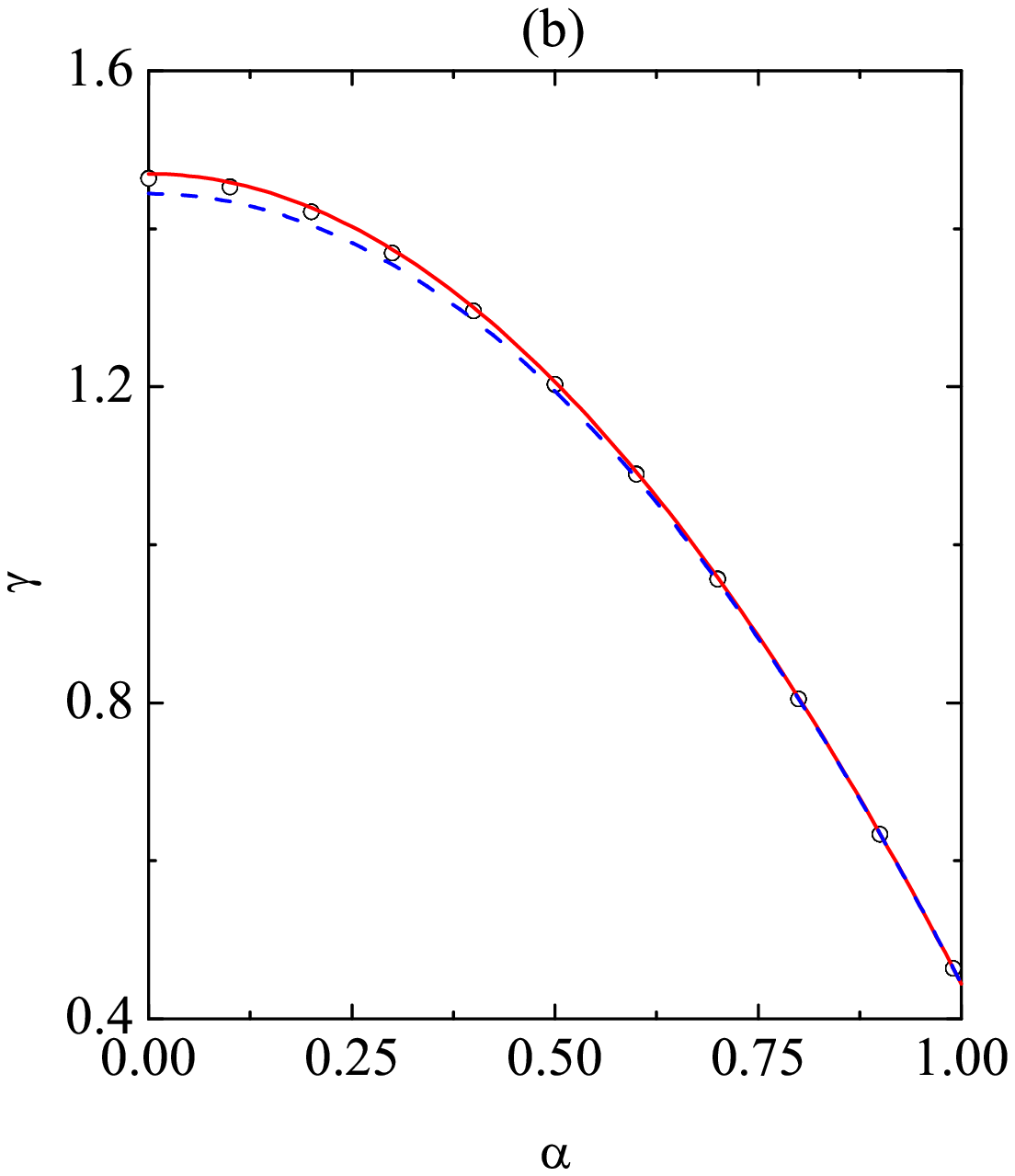}
\includegraphics[height=\widthtwocm]{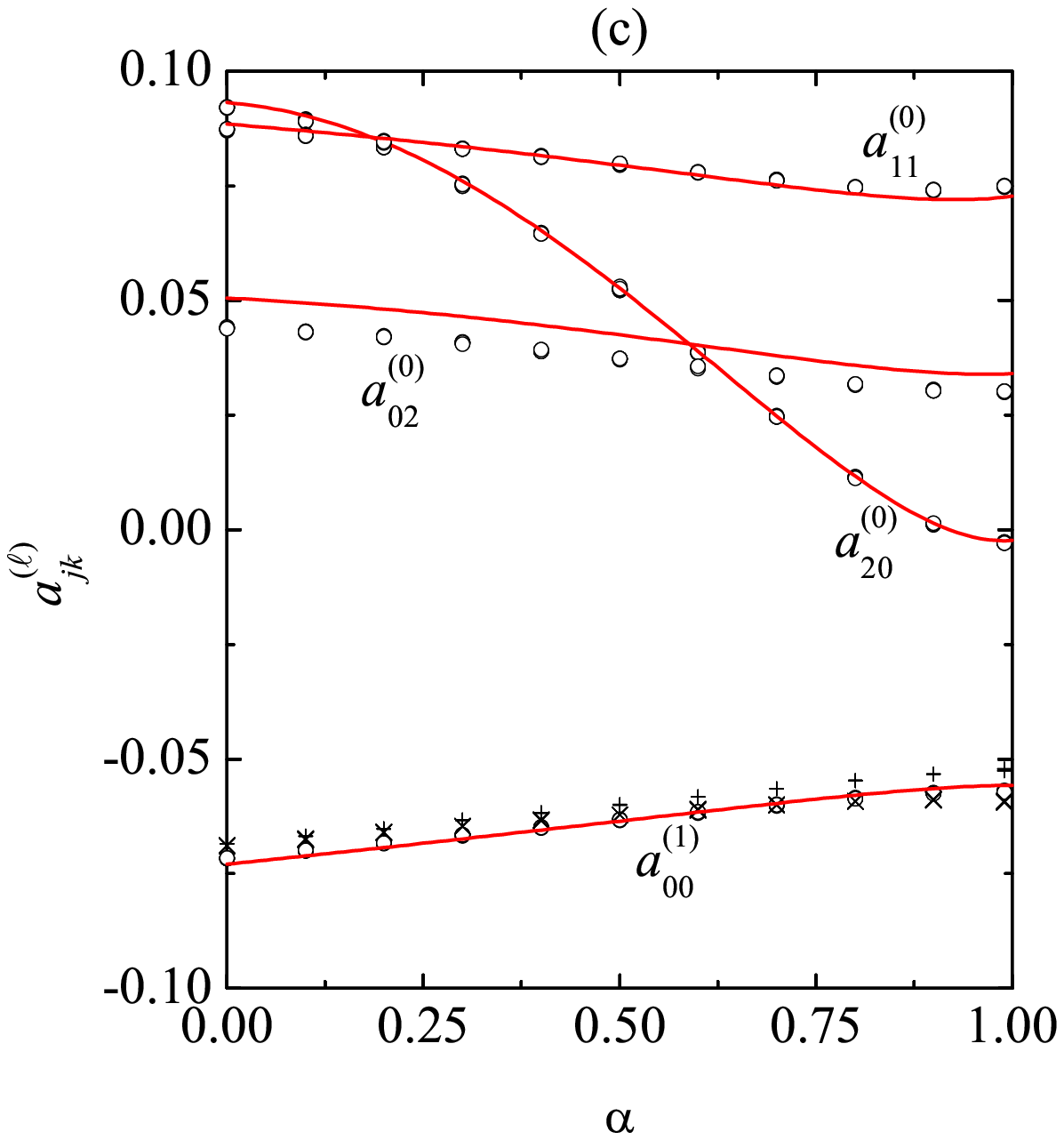}
\caption{
{Steady-state values of (a) the temperature ratio $\theta$, (b) the reduced noise intensity $\gamma$, and (c) the cumulants $a_{jk}^{(\ell)}$ as functions of $\alpha$ for $\beta=0$. Circles stand for DSMC data [with $\gamma(0)=\frac{3}{10}\sqrt{\pi},\frac{3}{5}\sqrt{\pi},\frac{6}{5}\sqrt{\pi},\frac{9}{5}\sqrt{\pi},3\sqrt{\pi},
\frac{18}{5}\sqrt{\pi}$], solid lines for the Sonine approximation, and dashed lines [only in panels (a) and (b)] for the Maxwellian approximation. The DSMC data for the quantities $b$ ($\times$) and $h$ ($+$) [see Eq.\ \eqref{eqb}] are also included in panel (c).}
\label{beta0}}
\end{figure}
\begin{figure}
\includegraphics[height=\widthtwocm]{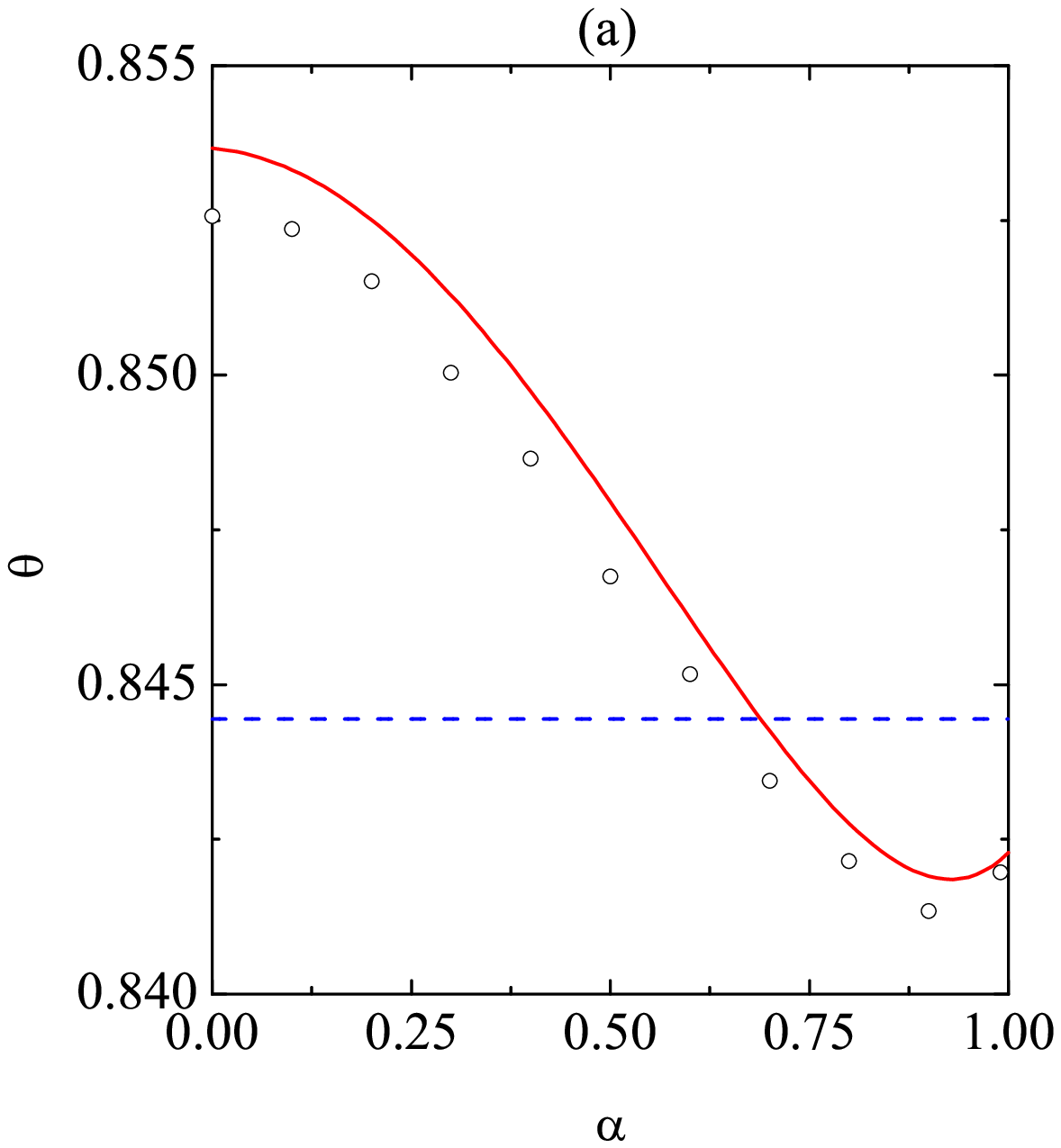}
\includegraphics[height=\widthtwocm]{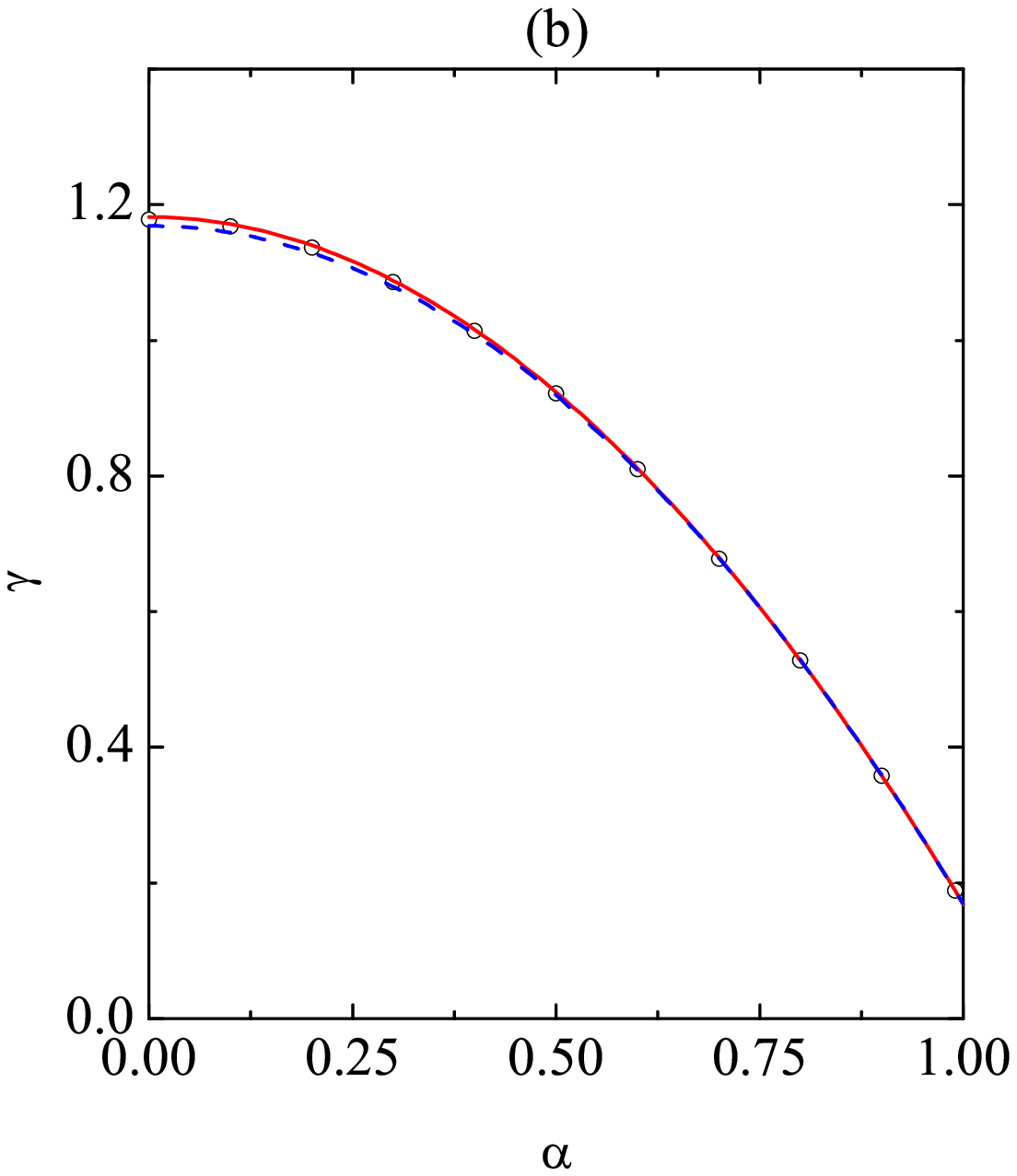}
\includegraphics[height=\widthtwocm]{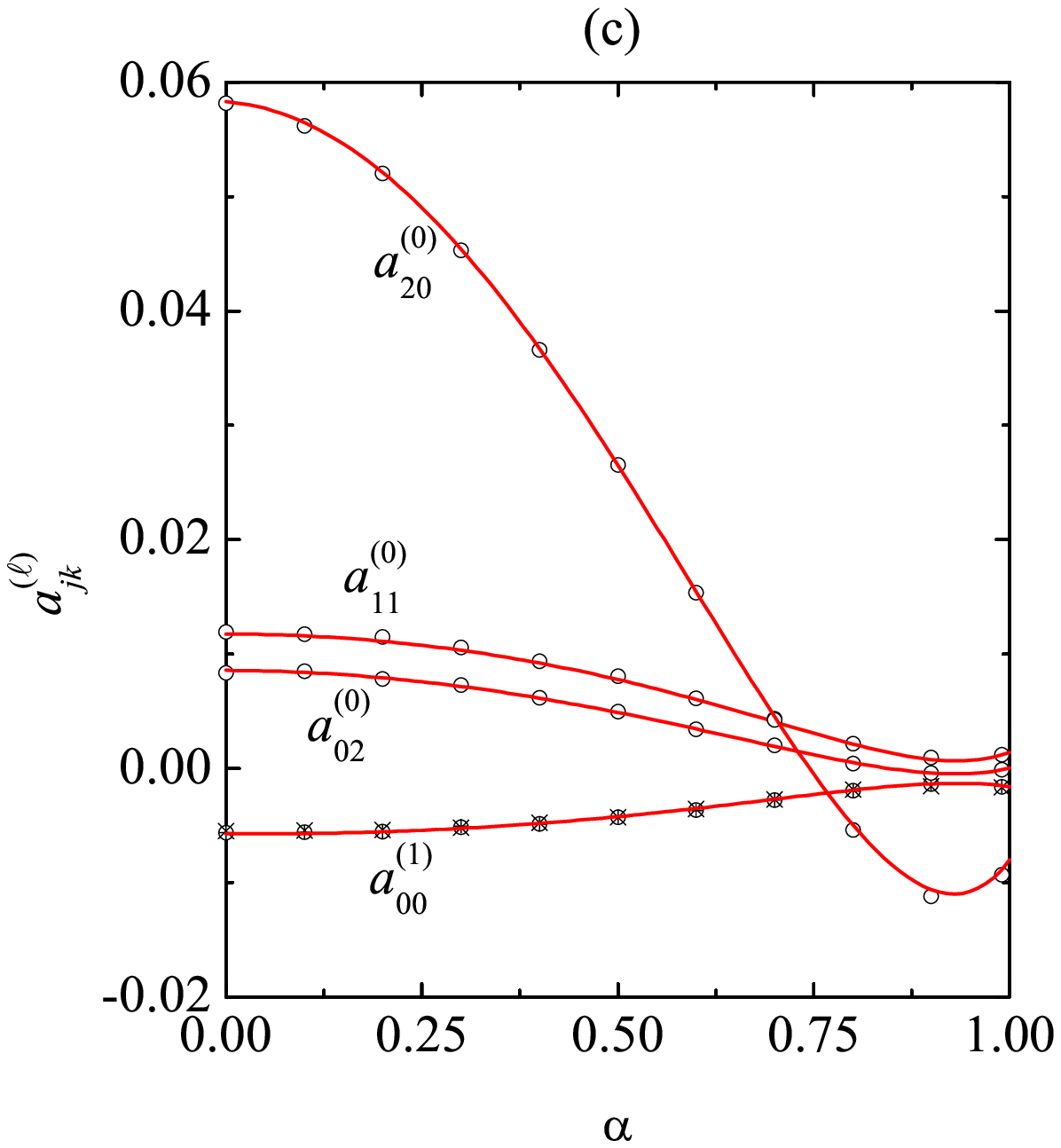}
\caption{
{Steady-state values of (a) the temperature ratio $\theta$, (b) the reduced noise intensity $\gamma$, and (c) the cumulants $a_{jk}^{(\ell)}$ as functions of $\alpha$ for $\beta=0.9$. Circles stand for DSMC data [with $\gamma(0)=\frac{18}{5}\sqrt{\pi}\simeq 6.381$], solid lines for the Sonine approximation, and dashed lines [only in panels (a) and (b)] for the Maxwellian approximation. The DSMC data for the quantities $b$ ($\times$) and $h$ ($+$) [see Eq.\ \eqref{eqb}] are also included in panel (c).}
\label{beta09}}
\end{figure}

{Now we turn to the analysis of the influence of $\alpha$ at fixed $\beta$. To that end, we have chosen the characteristic values $\beta=-0.9$ (weak roughness), $\beta=0$ (medium roughness),  and $\beta=0.9$ (strong roughness). The results are displayed in Figs.\ \ref{betam09}, \ref{beta0}, and \ref{beta09}, respectively.
Note that in Fig.\ \ref{beta0} the plotted DSMC data correspond to six different values of the initial reduced noise intensity ranging from $\gamma(0)=\frac{3}{10}\sqrt{\pi}\simeq 0.532$ to $\gamma(0)=\frac{18}{5}\sqrt{\pi}\simeq 6.381$. The almost perfect collapse of the points confirms the independence of the steady state on the values of $\chi_0^2$ and, equivalently, $\gamma(0)$.}

{From Figs.\ \ref{betam09}(a), \ref{beta0}(a), and \ref{beta09}(a) we observe that, even though the Maxwellian approximation states that the temperature ratio $\theta$ is independent of $\alpha$, this quantity tends to increase with increasing inelasticity, except near the elastic limit ($\alpha\lesssim 1$) for $\beta=-0.9$ and $0.9$, were a minimum value is reached. These fine-grained phenomena are very well accounted for by our Sonine approximation, although the latter tends to slightly overestimate $\theta$ (note the magnified vertical scales). At fixed $\beta$ a decrease of $\alpha$ produces an increase of dissipation and, consequently, an increase of $\gamma$. According to Figs.\ \ref{betam09}(b), \ref{beta0}(b), and \ref{beta09}(b), the parabolic dependence of $\gamma$ on $\alpha$ predicted by the Maxwellian approximation [see Eq.\ \eqref{thetaMax}] tends to lie slightly below the DSMC points for small $\alpha$. This deviation is satisfactorily corrected by the Sonine approximation.
In what concerns the cumulants, Figs.\ \ref{betam09}(c), \ref{beta0}(c), and \ref{beta09}(c) show again a very good predictive power of the Sonine approximation. As in Fig.\ \ref{alfa09}(c), the larger discrepancies are found for $a_{02}^{(0)}$ for medium roughness ($\beta=0$). Interestingly, the translational kurtosis  becomes larger than the rotational one, i.e., $a_{20}^{(0)}>a_{20}^{(0)}$ if $\alpha$ is small enough. Also, the practical equivalence between the correlation quantities $a_{00}^{(1)}$, $b$, and $h$ is maintained for all $\alpha$, although small deviations can be observed at $\beta=-0.9$ and $\beta=0$.}

\subsection{Marginal velocity distribution functions}

\begin{figure}
\includegraphics[height=\widthone]{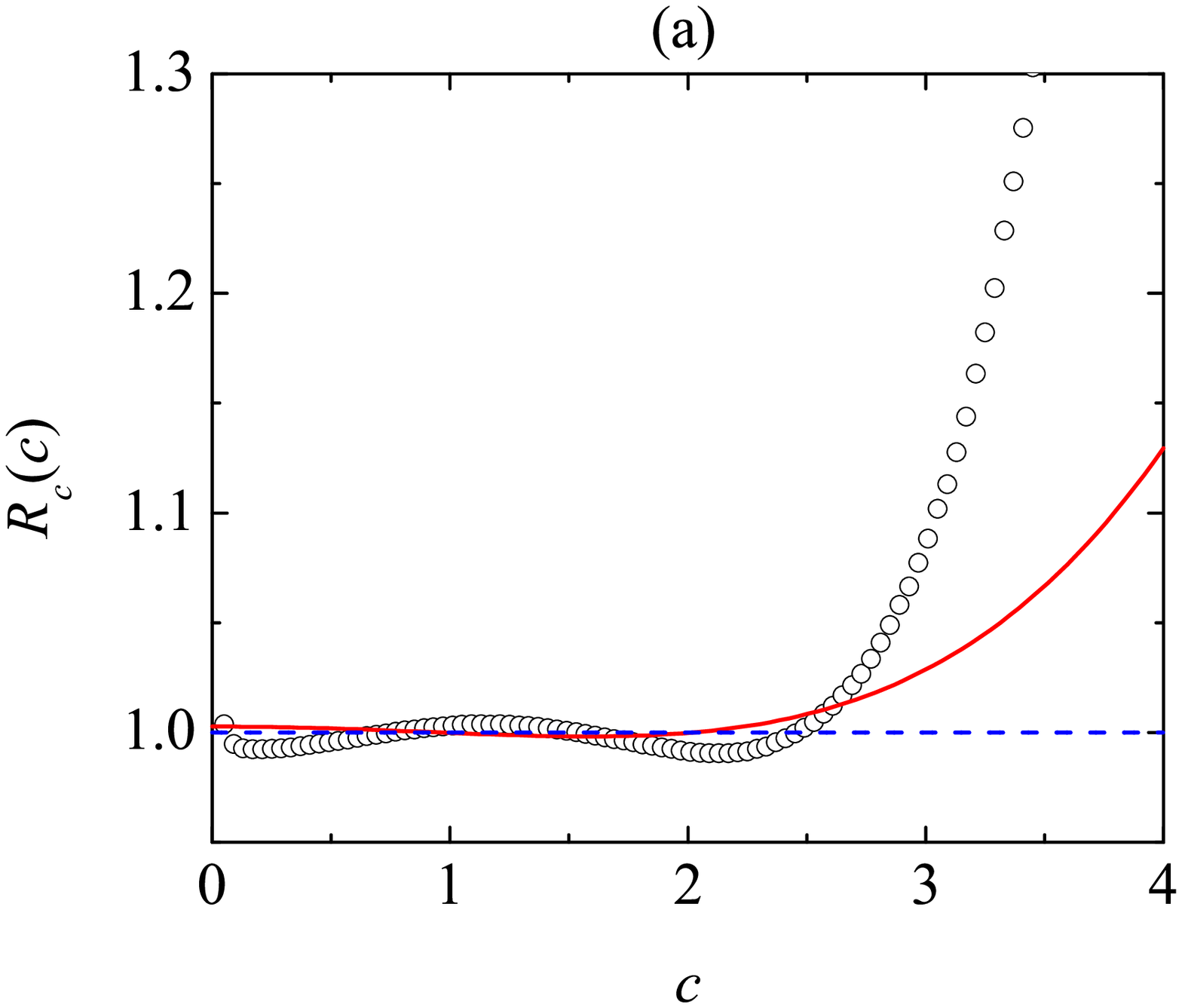}
\includegraphics[height=\widthone]{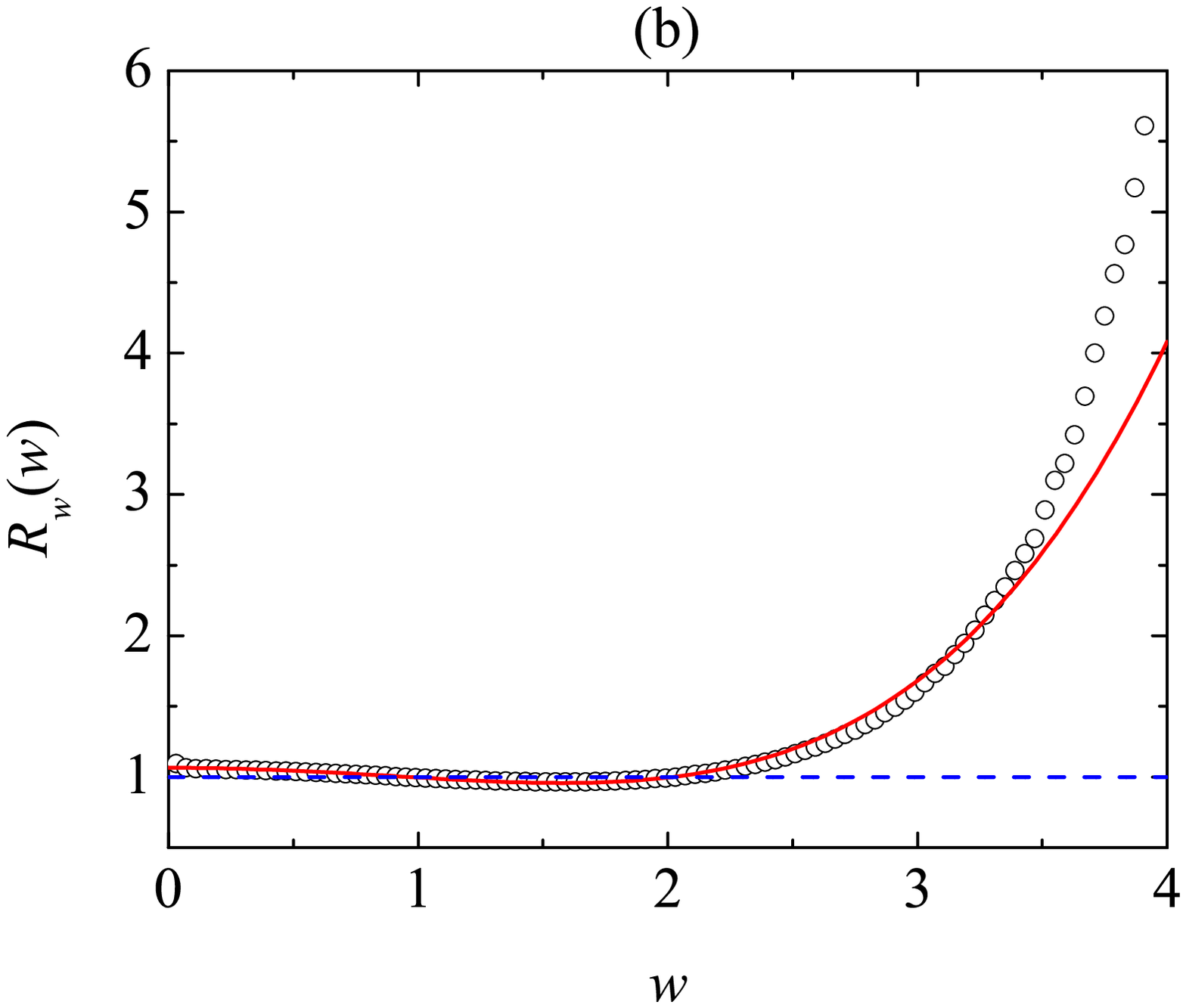}\\
\vspace{0.5cm}
\includegraphics[height=\widththree]{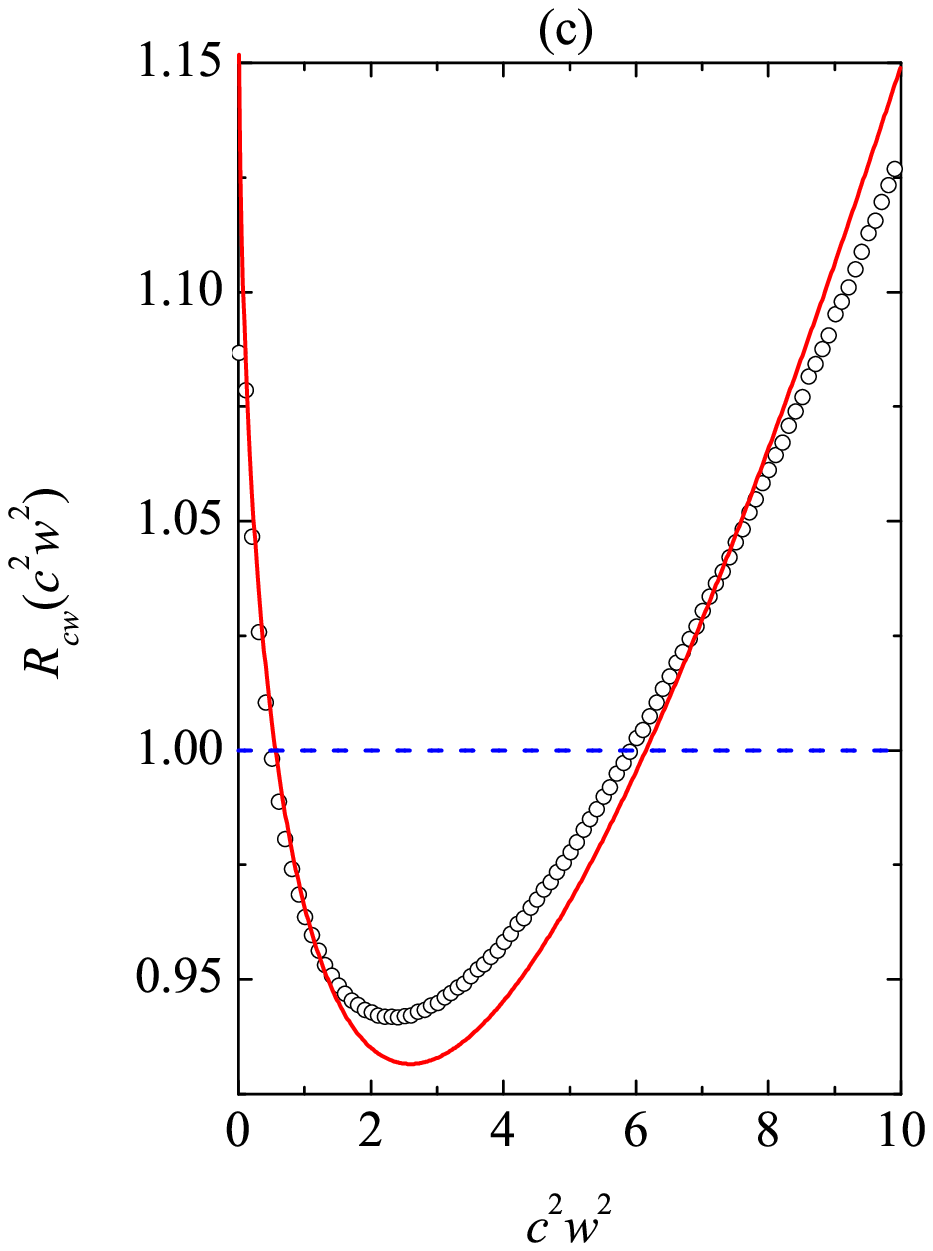}
\includegraphics[height=\widththree]{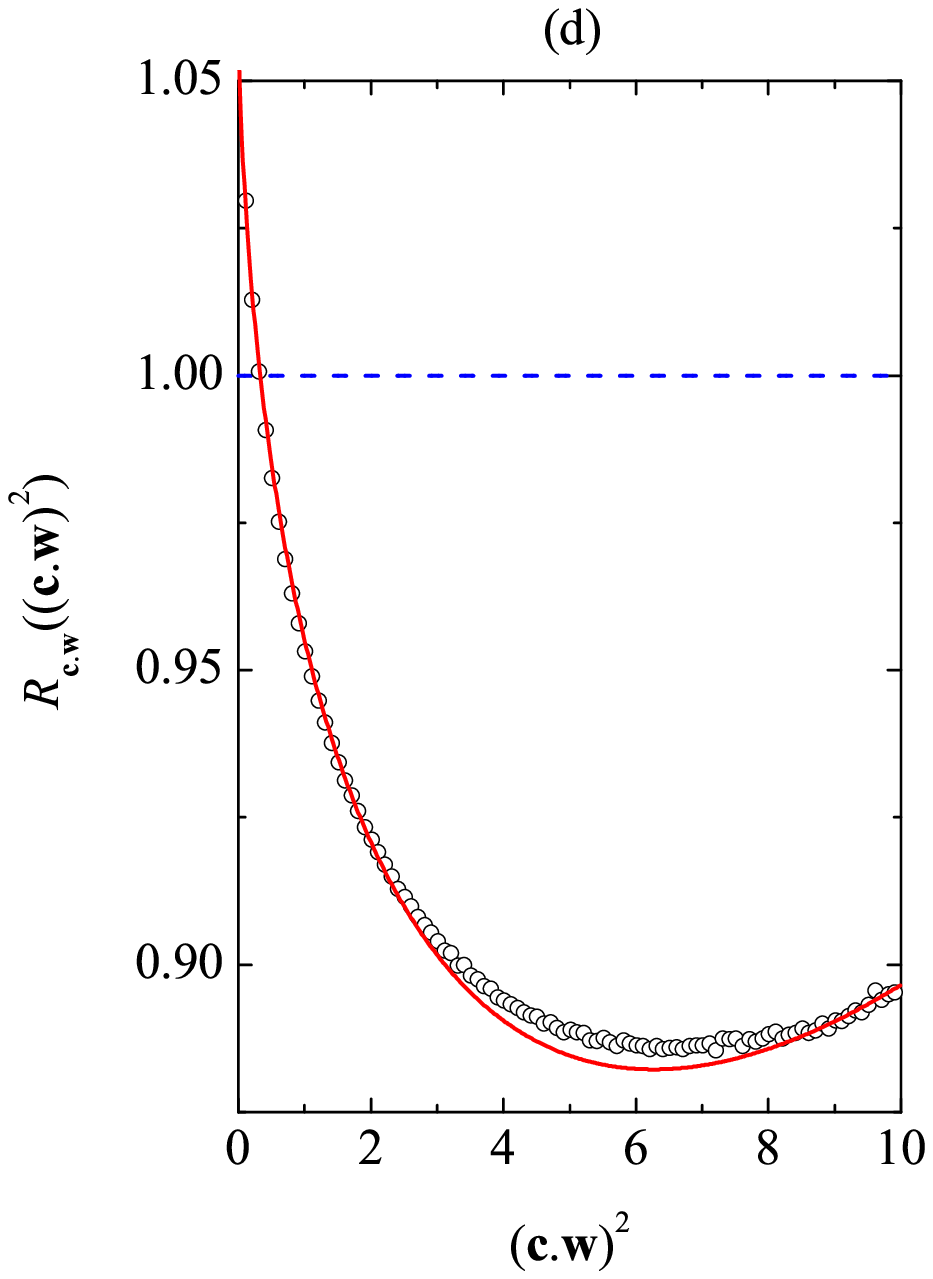}
\includegraphics[height=\widththree]{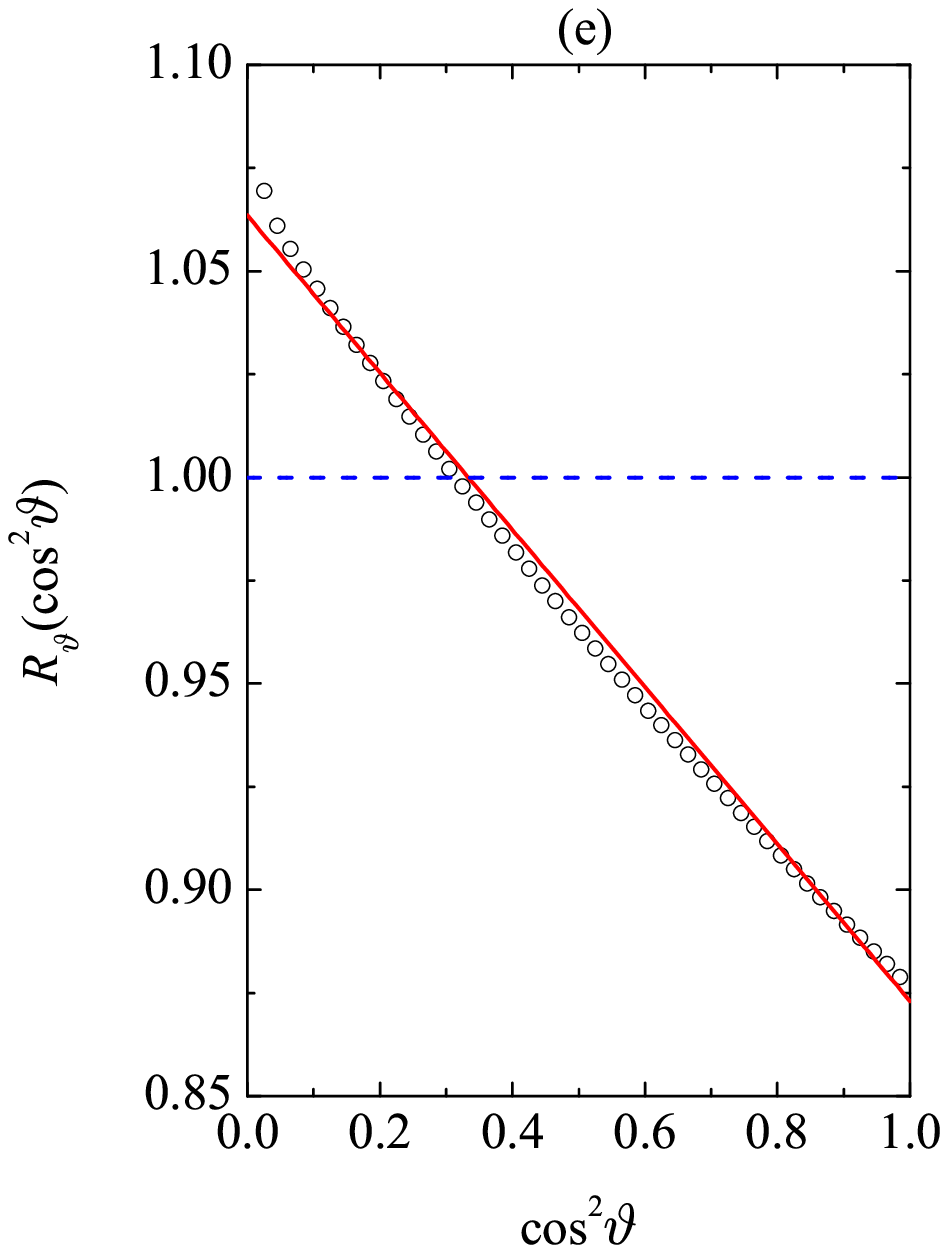}
\caption{Ratios of marginal distribution functions (a) $R_c(c)=\phi_c(c)/\phi_{c,M}(c)$, (b) $R_w(w)=\phi_w(w)/\phi_{w,M}(w)$, (c) $R_{cw}(c^2w^2)=\phi_{cw}(c^2w^2)/\phi_{cw,M}(c^2w^2)$, (d) $R_{\mathbf{c}\cdot\mathbf{w}}((\mathbf{c}\cdot\mathbf{w})^2)=
\phi_{\mathbf{c}\cdot\mathbf{w}}((\mathbf{c}\cdot\mathbf{w})^2)/\phi_{\mathbf{c}\cdot\mathbf{w},M}((\mathbf{c}\cdot\mathbf{w})^2)$, and (e) $R_{\vartheta}(\cos^2\vartheta)=\phi_{\vartheta}(\cos^2\vartheta)/\phi_{\vartheta,M}(\cos^2\vartheta)$ for $\alpha=0.9$ and $\beta=0$. Symbols stand for DSMC results and lines for the Sonine approximation theory [see Eqs.\ \protect\eqref{marg_Sonine}].
\label{marginal}}
\end{figure}

The four fourth-degree cumulants $a_{20}^{(0)}$, $a_{02}^{(0)}$, $a_{11}^{(0)}$, and $a_{00}^{(1)}$ encapsulate the basic piece of information about the deviations of the true velocity distribution function $\phi(\mathbf{c},\mathbf{w})$ from the (two-temperature) Maxwellian $\phi_M(c,w)=\pi^{-3}\exp(-c^2-w^2)$. On the other hand, it seems interesting to perform a more direct comparison between $\phi(\mathbf{c},\mathbf{w})$ and $\phi_M(c,w)$ and assess to what extent $\phi(\mathbf{c},\mathbf{w})$ can be represented by the truncated Sonine expansion \eqref{phi2}. Given that $\phi(\mathbf{c},\mathbf{w})$ depends on the three quantities $c^2$, $w^2$, and $(\mathbf{c}\cdot\mathbf{w})^2$ (or, analogously, $\cos^2\vartheta$), a plot of $\phi(\mathbf{c},\mathbf{w})$, or of the ratio $\phi(\mathbf{c},\mathbf{w})/\phi_M(c,w)$, would require a four-dimensional space. Therefore, it is convenient to analyze instead the projections of $\phi(\mathbf{c},\mathbf{w})$ represented by the five marginal distributions \eqref{6.1+6.2+6.5}.

Figure \ref{marginal} displays the ratios $R_c(c)=\phi_c(c)/\phi_{c,M}(c)$,  $R_w(w)=\phi_w(w)/\phi_{w,M}(w)$, $R_{cw}(c^2w^2)=\phi_{cw}(c^2w^2)/\phi_{cw,M}(c^2w^2)$,  $R_{\mathbf{c}\cdot\mathbf{w}}((\mathbf{c}\cdot\mathbf{w})^2)=
\phi_{\mathbf{c}\cdot\mathbf{w}}((\mathbf{c}\cdot\mathbf{w})^2)/\phi_{\mathbf{c}\cdot\mathbf{w},M}((\mathbf{c}\cdot\mathbf{w})^2)$, and  $R_{\vartheta}(\cos^2\vartheta)=\phi_{\vartheta}(\cos^2\vartheta)/\phi_{\vartheta,M}(\cos^2\vartheta)$ for the representative system $\alpha=0.9$, $\beta=0$. As we saw from Fig.\ \ref{alfa09}(c), at this value of the coefficient of normal restitution $\beta$ the magnitudes of all the cumulants, except the translational kurtosis $a_{20}^{(0)}$ are close to their maxima for $\alpha=0.9$. As can be observed from Fig.\ \ref{marginal}(a), although the Sonine approximation predicts very accurately the small value of $a_{20}^{(0)}$ ($a_{20}^{(0)}=0.00144$ versus the DSMC value $0.00137$), the function $R_c(c)$ is only qualitatively captured by Eq.\ \eqref{R_c}. In particular, the high-velocity tail of $R_c(c)$ (for $c\gtrsim 2.5$) is higher than predicted by the Sonine approximation. In contrast, even though the Sonine approximation overestimates the rotational kurtosis by about 12\% ($a_{02}^{(0)}=0.0343$ versus the DSMC value $0.0305$), Fig.\ \ref{marginal}(b) shows that the predicted function $R_w(w)$ is rather accurate up to $w\simeq 3.4$, where $R_w(w)$ reaches values as high as $2.5$. Again, however, the high-velocity tail of $R_w(w)$ is more overpopulated than the Sonine prediction. In fact, the high-velocity tails of the velocity distribution, which typically do not influence the first few cumulants, are not expected to be accounted for by any truncated Sonine expansion.

Now, we turn to the three marginal distribution functions related to the translational-rotational correlations. Figure \ref{marginal}(c) shows that small ($c^2w^2\lesssim 0.56$) and large ($c^2w^2\gtrsim 6.15$) values of $c^2w^2$ are more probable than it might be expected from the Maxwellian distribution. This phenomenon is well described by the Sonine approximation \eqref{R_cw}, although some differences are observed around the minimum of $R_{cw}(c^2w^2)$. Anyway, the error in $\langle c^2w^2\rangle-\frac{9}{4}$ is less than 3\% ($\frac{9}{4}a_{11}^{(0)}=0.162$ versus the DSMC value $0.167$).
According to Fig.\  \ref{marginal}(d), most of the values of $(\mathbf{c}\cdot\mathbf{w})^2$ within the ``thermal'' range are less probable than the Maxwellian expectations. This agrees with a negative value of $\langle(\mathbf{c}\cdot\mathbf{w})^2\rangle-\frac{3}{4}$, which is very well described by the Sonine approximation  ($\frac{3}{4}a_{11}^{(0)}+\frac{15}{8}a_{00}^{(11)}=-0.0518$ versus the DSMC value $-0.0520$). Note that the positive value of $a_{11}^{(0)}$ (i.e., $\langle c^2 w^2\rangle>\langle c^2\rangle\langle w^2\rangle$) is dominated  by a negative value of $\frac{5}{2}a_{00}^{(1)}$ (i.e., $3\langle (\mathbf{c}\cdot\mathbf{w})^2\rangle<\langle c^2 w^2\rangle$), resulting in $3\langle (\mathbf{c}\cdot\mathbf{w})^2\rangle<\langle c^2\rangle\langle w^2\rangle<\langle c^2 w^2\rangle$. Obviously, in order to preserve normalization, $R_{\mathbf{c}\cdot\mathbf{w}}((\mathbf{c}\cdot\mathbf{w})^2)$ must be larger than unity for values of $(\mathbf{c}\cdot\mathbf{w})^2$ higher than those considered in our DSMC simulations (according to the Sonine approximation, this would happen for $(\mathbf{c}\cdot\mathbf{w})^2>19.26$), but they do not affect the sign of $\langle(\mathbf{c}\cdot\mathbf{w})^2\rangle-\frac{3}{4}$.
Finally, Fig.\ \ref{marginal}(e) confirms the lifted-tennis-ball effect, i.e., the values of $\cos^2\vartheta$ smaller than about $\frac{1}{3}$ are more probable than one could expect from a Maxwellian distribution, so that $\langle \cos^2\vartheta\rangle-\frac{1}{3}=\frac{3}{10}b<0$. The Sonine approximation predicts the former difference within an error smaller than  5\% ($\frac{3}{10}b=-0.0169$ versus the DSMC value $-0.0177$). Apart from the average value  $\langle \cos^2\vartheta\rangle$, the Sonine approximation \eqref{R_theta} describes rather accurately the whole distribution $R_\vartheta(\cos^2\vartheta)$, as  Fig.\ \ref{marginal}(e) shows.

\section{Conclusion}
\label{sec5}

We have developed in this work a kinetic theory for a dilute granular gas of rough hard spheres that is heated by a uniform stochastic volume force. Our theory is based on a {Sonine} polynomial expansion of the granular gas distribution function around the Maxwellian. Thus, we obtain a numerical solution of the velocity distribution function at all times and an analytical solution for the uniform steady state that is reached after a relatively short relaxation time. We have tested our theoretical approach against DSMC data, obtaining a very good agreement between simulation and theory.

Contrary to the results for a freely cooling granular gas of rough spheres in a recent work,\cite{VSK14} the cumulants of the distribution function  reach low values (always below $0.1$, as measured from simulations and theory) and thus our theoretical approach works in the whole range of inelasticity and roughness values. In other words, it would not be necessary for the heated granular gas to add higher order terms in the truncated Sonine expansion that we have considered, as  done in recent works on the HCS for inelastic rough spheres.\cite{RA14} Also, thanks to the existence of a true steady state (where inelastic cooling and external heating balance each other), the quasismooth limit $\beta\to -1$ is regular and thus the results for  purely smooth spheres ($\beta=-1$) are recovered in that limit. Again, this differs from the situation in the HCS.\cite{BPKZ07,KBPZ09,VSK14}

Another relevant result for the heated granular gas is that the translations and rotations in particles, although still correlated as in the HCS,\cite{BPKZ07} show a very mince cannon-ball effect region and this effect completely disappears from the nearly-smooth region, appearing only  near the completely elastic and rough limit. This is in contrast with the behavior in the HCS, where two neat cannon-ball regions may be found.\cite{BPKZ07,KBPZ09,RA14}

Summarizing, the white noise force has an effect of making the distribution function more similar to the Maxwellian for the whole range of coefficients of normal and tangential restitution. Since our Sonine expansion is done around the Maxwellian we expect the theoretical solution that we provide in this work to be an accurate description and a reference for researchers interested in the description of the transport properties of a heated granular gas of rough spheres.

\begin{acknowledgments}
The authors acknowledge support from the Spanish Government through Grant No.\ FIS2013-42840-P and from the
Regional Government of Extremadura (Spain) through Grant No.\ GR15104 (partially financed by the European Regional Development Fund, ERDF).
Computing facilities from Extremadura Research Centre for Advanced Technologies (CETA-CIEMAT), funded by the ERDF, are gratefully acknowledged.
\end{acknowledgments}

\appendix*
\section{Collisional moments}
\label{collmoments_appendix}

We write in this Appendix the {Sonine-approximation} expressions of the collisional moments $\mu_{pq}^{(2r)}$ {(with $p+q+2r=2$ and $4$)} as {linear} functions of the cumulants $a_{jk}^{(\ell)}$ {(with $j+k+2\ell=2$) and nonlinear functions of the temperature ratio $\theta$}. Those expressions, where we we make use of the parameters \eqref{7}, are
\beq
{\mu_{20}^{(0)}}={4\left[\at\left(1-\at\right)+\bt\left(1-\bt\right)\right]
\left(1+\frac{3a_{20}^{(0)}}{16}\right)}{-\frac{4\bt^2\theta}{\kappa}
\left(1-\frac{a_{20}^{(0)}}{16}+\frac{2a_{11}^{(0)}{-a_{00}^{(1)}}}{8}\right)},
\label{22}
\eeq
\beq
\mu_{02}^{(0)}=\frac{4\bt}{\kappa}\left[\left(1-\frac{\bt}{\kappa}\right)\left(1-\frac{a_{20}^{(0)}}{16}+
\frac{2a_{11}^{(0)}{-a_{00}^{(1)}}}{8}\right)-\frac{\bt}{\theta}\left(1+\frac{3a_{20}^{(0)}}{16}\right)\right],
\label{23}
\eeq
\beqa
\mu_{40}^{(0)}&=&16\left[\at^3\left(2-\at\right)+\bt^3\left(2-\bt\right)-\at\bt \left(1-\at-\bt+\at\bt\right)\right]+22\left(\at+\bt\right)\nn
&&-38\left(\at^2+\bt^2\right)-15\Bigg[\at\bt \left(\frac{23}{15}-\at-\bt+\at\bt\right)-\frac{269}{120}\left(\at+\bt\right)+\frac{357}{120}\left(\at^2+\bt^2\right)
\nn
&&-\at^3\left(2-\at\right)-\bt^3\left(2-\bt\right)\Bigg]
a_{20}^{(0)}-\frac{22 \bt^2\theta}{\kappa}\left(1+\frac{41a_{20}^{(0)}}{176 }+3\frac{2a_{11}^{(0)}{-a_{00}^{(1)}}}{8}\right)\nn
&&+\frac{16\bt^2\theta}{\kappa}\left[\at\left(1-\at\right)+2\bt\left(1-\bt\right)\right]
\left(1+\frac{3a_{20}^{(0)}}{16}+3\frac{2a_{11}^{(0)}{-a_{00}^{(1)}}}{8}\right)\nn
&&-\frac{16\bt^4 \theta^2}{\kappa^2}\left(1-\frac{a_{20}^{(0)}}{16}+\frac{a_{02}^{(0)}}{2}+\frac{2a_{11}^{(0)}{-a_{00}^{(1)}}}{4}\right),
\label{28}
\eeqa
\beqa
\mu_{22}^{(0)}&=&
6\left[\at\left(1-\at\right)+\bt\left(1-\bt\right)-\frac{4\at\bt}{3\kappa}\left(1-\at\right)\left(1-\frac{\bt}{\kappa}\right)-\frac{8\bt^2}{3\kappa}
\left(\frac{3}{4}-\bt-\frac{\bt}{\kappa}+2\frac{\bt^2}{\kappa}\right)\right]
\nn
&&\times \left(1+\frac{3a_{20}^{(0)}}{16}+3\frac{2a_{11}^{(0)}{-a_{00}^{(1)}}}{8}\right)+\frac{7\bt}{\kappa}
\left(1-\frac{\bt}{\kappa}\right)\left(1+\frac{29a_{20}^{(0)}}{112}\right)-\frac{3\bt^2}{2\kappa\theta}a_{20}^{(0)}
\nn
&&-\frac{8\bt^2}{\kappa\theta}\left[\frac{9}{8}-\at\left(1-\at\right)-2\bt\left(1-\bt\right)
\right] \left(1+\frac{15a_{20}^{(0)}}{16}\right)-\frac{\bt^2\theta}{\kappa}\left[5-8 \frac{\bt}{\kappa}\left(1-\frac{\bt}{\kappa}\right)\right]a_{02}^{(0)}\nn
&&-\frac{8\bt^2\theta}{\kappa}\left[1-2 \frac{\bt}{\kappa}\left(1-\frac{\bt}{\kappa}\right)\right] \left(1-\frac{a_{20}^{(0)}}{16}+\frac{2a_{11}^{(0)}{-a_{00}^{(1)}}}{4}\right)+3\Bigg[\frac{\bt}{\kappa}
\left(\frac{37}{12}-2\bt-\frac{7\bt}{4\kappa}\right)\nn
&&+\at+\bt-\frac{4 \at\bt}{3\kappa}\Bigg]\frac{2a_{11}^{(0)}{-a_{00}^{(1)}}}{2}{+\Bigg[5\left(\at+\bt\right)-3\left(\at^2+\bt^2\right)}
{+\frac{4\bt}{\q}\left(1-\bt\right)}\nn
&&{-\frac{\bt^2}{\q^2}\left(2+\q\theta\right)\Bigg] \frac{3a_{00}^{(1)}}{4}},
\label{29}
\eeqa
\beqa
\mu_{04}^{(0)}&=&\frac{\bt}{\kappa}\left\{4\left(1-\frac{\bt}{\kappa}\right)\left[5-4\frac{\bt}{\kappa}
\left(1-\frac{\bt}{\kappa}\right)\right]\left(1-\frac{a_{20}^{(0)}}{16}\right)- \frac{4\bt}{\theta}\left[5-8\frac{\bt}{\kappa}\left(1-\frac{\bt}{\kappa}\right)\right]
\right.\nn
&&\times \left(1+\frac{3a_{20}^{(0)}}{16}+3\frac{2a_{11}^{(0)}{-a_{00}^{(1)}}}{8}\right)-{5}\left(1-\frac{4 \bt}{5\kappa}\right)\left({2a_{11}^{(0)}{-a_{00}^{(1)}}}\right)-\frac{16\bt^3}{\kappa\theta^2}\left(1+\frac{15a_{20}^{(0)}}{16}\right)
\nn
&&\left.
+4
\left(5-\frac{13}{2}\frac{\bt}{\kappa}+4\frac{\bt^2}{\kappa^2} - 2\frac{\bt^3}{\kappa^3}\right)\left(a_{02}^{(0)}+\frac{2a_{11}^{(0)}{-a_{00}^{(1)}}}{2}\right)
{+\left(1-\frac{\bt}{\kappa}-\frac{\bt}{\theta}\right)\frac{3a_{00}^{(1)}}{2}}\right\},
\label{30}
\eeqa
\beqa
{\mu_{00}^{(2)}}&=&2\left[\at\left(1-\at\right)-\bt^2\left(1+\frac{1}{\q^2}\right)\right]
\left(1+\frac{3a_{20}^{(0)}}{16}+\frac{3a_{11}^{(0)}}{4}+\frac{3a_{00}^{(1)}}{4}\right)
+{\at}\left(a_{11}^{(0)}+4a_{00}^{(1)}\right)
\nn
&&+2\bt\left(1+\frac{1-\bt}{\q}\right)\left(1+\frac{3a_{20}^{(0)}}{16}+\frac{5a_{11}^{(0)}}{4}+\frac{13a_{00}^{(1)}}{8}\right)
+3\bt\left(\frac{3}{4}-\at\right)\left(1+\frac{1}{\q}\right){a_{00}^{(1)}}
\nn
&&-\frac{\bt^2}{\q\theta}\left(1+\frac{7a_{20}^{(0)}}{16}\right)-\frac{\bt^2\theta}{\q}\left(1-\frac{a_{20}^{(0)}}{16}+\frac{2a_{11}^{(0)}{-a_{00}^{(1)}}}{4}\right).
\label{30n}
\eeqa


%

\end{document}